\documentclass[12pt,aps,reprint,superscriptaddress,a4paper]{revtex4-2}
\usepackage{graphicx}
\usepackage{xcolor}
\usepackage{subcaption}
\begin{document}

\title{Impurity atom configurations in diamond and their visibility via scanning transmission electron microscopy imaging}

\author{D. Propst}
\affiliation{University of Vienna, Faculty of Physics, Boltzmanngasse 5, 1090 Vienna, Austria}
\affiliation{University of Vienna, Vienna Doctoral School in Physics, Boltzmanngasse 5, 1090 Vienna, Austria}
\author{J. Kotakoski}
\email{jani.kotakoski@univie.ac.at}
\affiliation{University of Vienna, Faculty of Physics, Boltzmanngasse 5, 1090 Vienna, Austria}
\author{E. H. Åhlgren}\email{harriet.ahlgren@univie.ac.at}
\affiliation{University of Vienna, Faculty of Physics, Boltzmanngasse 5, 1090 Vienna, Austria}

\begin{abstract}
Dispersed impurities in diamond present a flourishing platform for research in quantum informatics, spintronics and single phonon emitters.
Based on the vast pool of experimental and theoretical work describing impurity atoms in diamond, we review the configurations by the chemical element discussing the relevant atomic configurations and most important properties.
Dopant structures expand from single to co-doping configurations, also combined with carbon vacancies.
Despite of their importance, not much is known about the exact atomic configurations associated with the dopant structures beyond computational models, partially due to difficulties in their microscopic observation.
To assess the visibility of these structures, we carry out image simulations to show that the heavier dopants may be easily discernible in scanning transmission electron microscopy annular dark field images, with a window of visibility of up to over $\pm$~10~nm in defocus.
We further present the first atomic resolution images of an impurity atom configuration (substitutional Er atom) in the diamond lattice, 
confirmed by a comparison to the simulated images.
Overall, our results demonstrate that there is a vast research field waiting for the microscopy community in resolving the exact atomic structure of various impurity atom configurations in diamond.
\end{abstract}

\maketitle

\vspace{2pc}

\tableofcontents

\section{Introduction}

Research on impurity atoms in diamond started as early as 1878~\cite{Endlich1878}.
Diamond is a wide band gap semiconductor ($5.5$~eV~\cite{clark1964_gap,CHENG2023_gap}). To gain control over diamond's electronic and optical properties, various dopants have been studied for their ability to introduce impurity states within its band gap.  
Impurities with the capability to introduce shallow donor and acceptor states have been intensively studied for electronic applications that require a material that can withstand high frequencies, high temperatures, high powers and high mechanical wear. 
A shallow acceptor is easily managed, with B-doped diamond showing $p-$type behavior~\cite{poferl-1973}, while the shallow donor state ($n$-type) has been more difficult to establish. 
More recently, the research has shifted towards optical properties since several dopant centers have been found optically active~\cite{Igor2014}. 
As a result, promising applications have arisen in the processing of quantum information, photonics, chemical analysis and localisation of molecules, as well as labeling and tracking, microfluidics, electrochemistry and sensing. 
Recent reviews have been published on the use of diamond color centers in chemical and biochemical analysis~\cite{Mzyk2022}, and photonics~\cite{Aharonovich2011,Igor2014,Schroder:16}. 
The optical properties were discussed in depth in the Handbook by Zaitsev in 2001~\cite{Zaitsev2001_handbook} and the ion implantation process was reviewed by T. Lühmann et al. in 2018~\cite{LühmannTobias2018Saeo}.

To date, the nitrogen-vacancy (N$V$) center is the most discussed impurity center in diamond.
The N$V$ center possesses many unparalleled properties that are realised in its ability to act as a solid-state quantum bit in ambient conditions, although not without drawbacks~\cite{Pezzagna_2011}.
Following the success of the N$V$ center, the group 14 elements, namely silicon-, germanium-, tin- and lead-vacancy centers have been under intense investigation~\cite{Bradac2019}. 
Recently, rare earth metals have been incorporated into diamond. This has been spurred on by their photoluminescent properties giving rise to sharp emission lines at varying wavelengths~\cite{JONES2006}. This could be used to create new  single photon emitters for the use in quantum optics. 

Meanwhile, imaging of individual impurity atoms inside semiconductor materials became possible only recently due to improved sample preparation techniques and instrumentation for (scanning) transmission electron microscopy ((S)TEM).
The first report was on individual antimony (Sb) impurity atoms in crystalline silicon, published by Voyles et al.
in 2002~\cite{Voyles2002}, but it was unable to pin down the exact atomic configuration associated with them.
More recently, nitrogen containing platelet defects were identified in natural type Ia diamond with TEM showing atomic resolution and the spectroscopic fingerprint~\cite{Olivier2018}.
Moreover, the spectroscopic identification of a single NV center was demonstrated in 2023~\cite{Hudak2023} in $4$~nm natural nanodiamond. 

Although direct imaging of individual impurity configurations inside the lattice would open a likely path to an unprecedented understanding of impurity configurations, their distribution and properties in diamond, there is a lack of studies aiming to directly image the impurity sites.
To motivate the community of microscopists to start working on this and to provide a comprehensive review on the structure and properties of impurity centers in diamond, we present guiding images of various impurity atom configurations studied with \textit{ab initio} structural relaxation and image simulations aimed for electron microscopy imaging.
We also provide the first STEM images of individual Er dopants implanted into diamond and imaged along the [110] axis. 

\section{Structure and properties of impurity atom configurations}

\begin{figure*}
\includegraphics[width=0.95\textwidth]{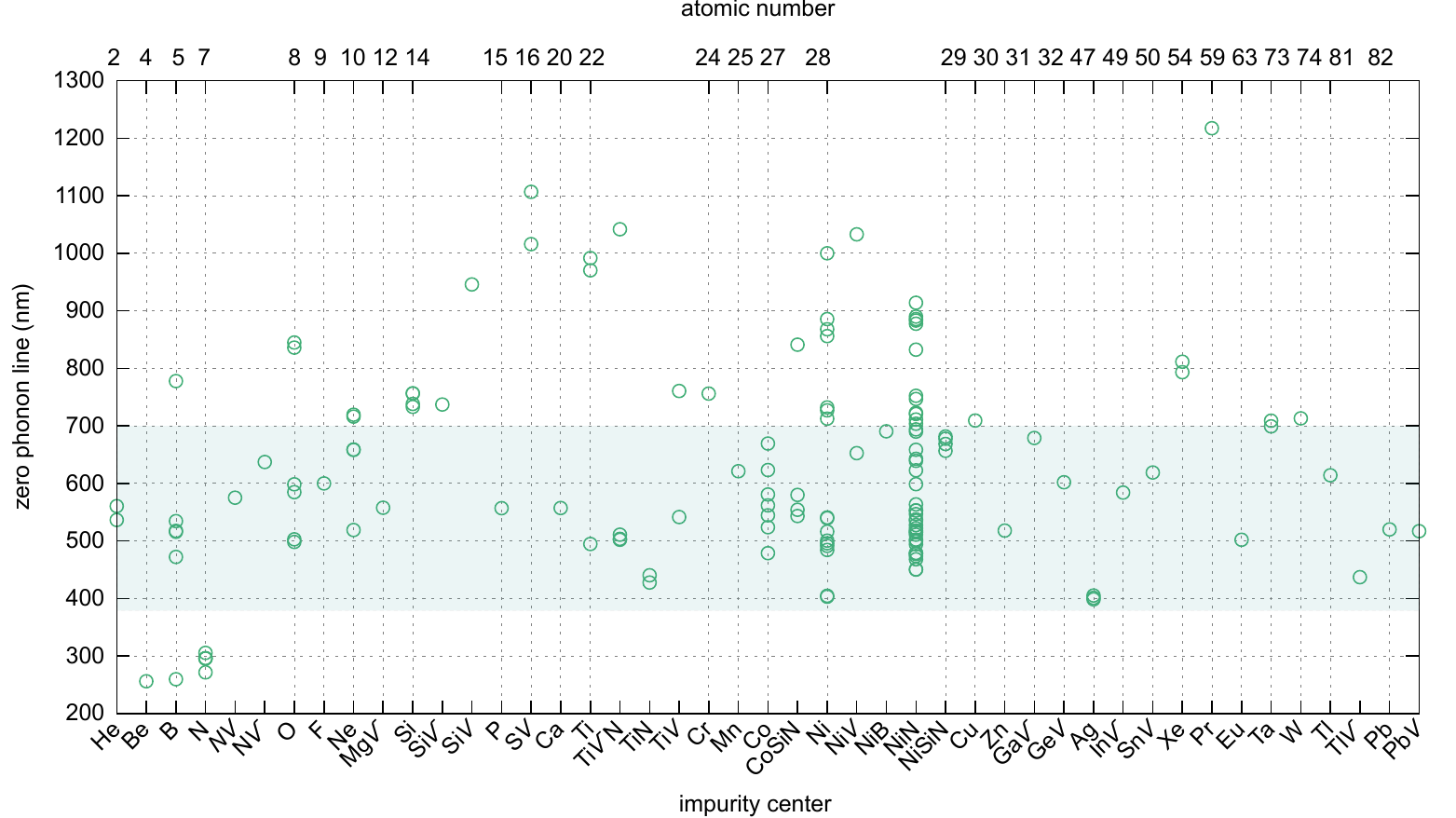}
    \caption {
        {\bf Optical transitions. }
        The zero phonon line describing the difference in energy levels between ground and excited states without the assistance of phonons for different impurity centers in diamond.
        Colored area (light green) indicates the range of the visible spectrum.
        $V$ in the name implies a vacancy, \textit{e.g.}, Ti$V^-$N refers to the negatively charged titanium plus a vacancy with a neighboring nitrogen atom.
        The values are cited in the text for each element.             
        }
\label{fg::zpl}
\end{figure*}

\begin{figure}
\includegraphics[width=0.45\textwidth]{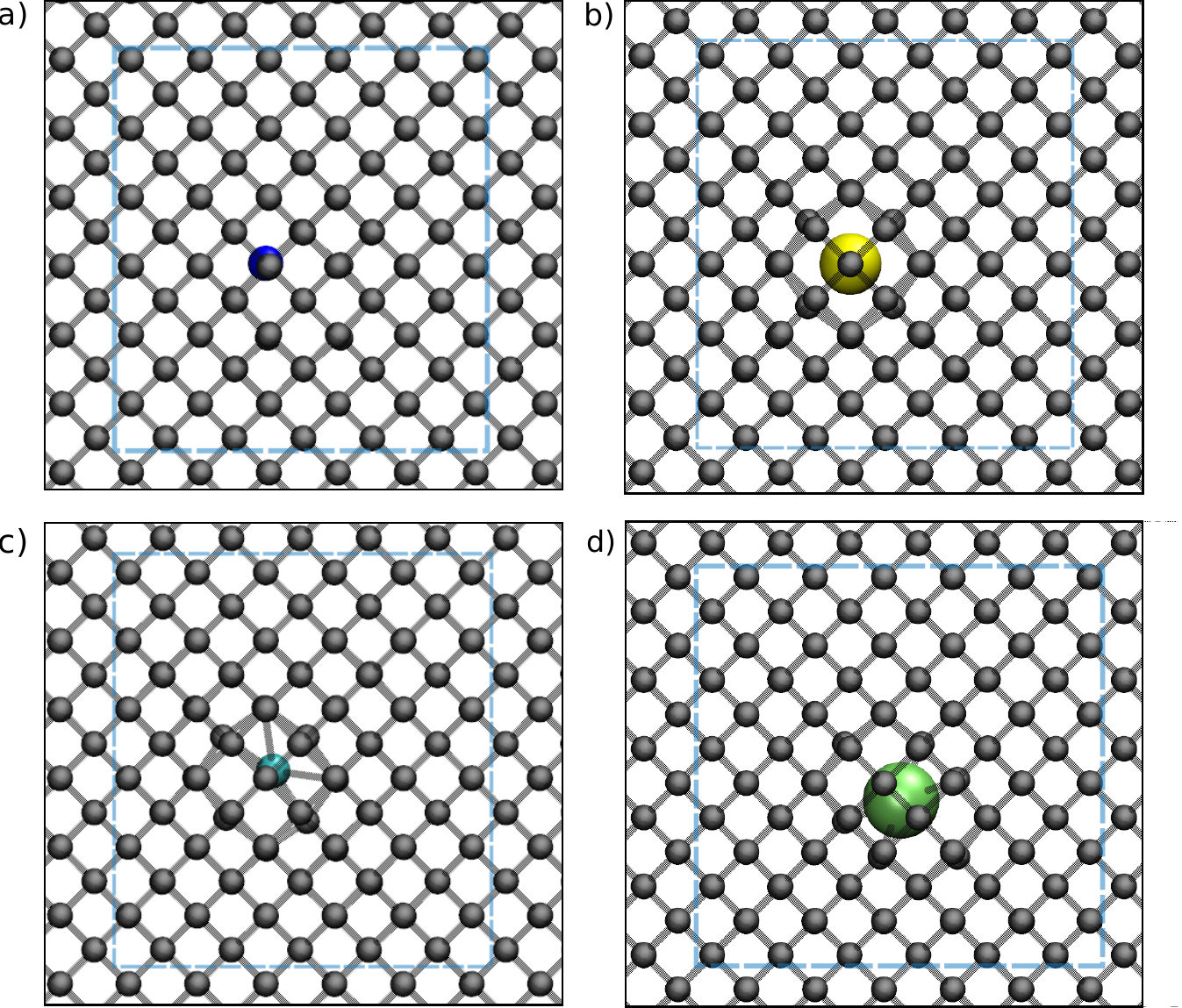}
    \caption {{\bf Structures of the DFT-relaxed dopant configurations in diamond, showing the general trends for different atomic species.}
    (a) A nitrogen with a vacancy (N$V$) with $C_{3v}$ symmetry with a shift along the [111] axis. (b) Antimony in a substitutional position. Its symmetry is close to $T_d$ but with a slight distortion in the surrounding lattice. (c) Chlorine substitution. (d) Erbium with a neighboring vacancy (Er$V$) showing the split-vacancy configuration found for mid to heavy element dopant-vacancy structures. The symmetry is close to $D_{3d}$.
    Dashed boundary indicates the size of the simulation cell.
    }
\label{fg:dft}
\end{figure}

In the following chapters we provide a short overview of impurity atom configurations that include a single dopant species, concentrating on their atomic structure and electronic/optical properties.
These configurations extend from a simple interstitial to more complex combinations with vacancies that are especially relevant for post-growth doping techniques of diamond, such as ion implantation.
Fig.~\ref{fg::zpl} shows the zero phonon lines for the discussed dopants, and in Fig.~\ref{fg:dft} the most common impurity atom structures are shown, as obtained from structural optimization with density functional theory (DFT), as explained later in section~\ref{sec::dft}.
Hydrogen-impurity complexes are added in the discussion when available research has been found, due to hydrogen's tendency to introduce impurity sites with compensating effects.
Other relevant co-doping complexes presented in the literature are discussed on the side.

\subsection{Group 1}

Group $1$ elements have been studied for their ability to create $n-$doped diamond.
Certain configurations show promise, although other sites act as compensating centers thus complicating the process.
Almost all complex co-doping configurations with hydrogen will result in deep donor sites.   

\begin{table}[h!]
    \caption{
    {\bf Group 1.} 
    States introduced in the diamond band gap by the group $1$ impurity centers are indicated by the nature of the doping: acceptor ($p$-type) and donor ($n$-type).
    }
    \begin{tabular}{ |c|c|c| }
    \hline
    impurity & gap state & zero phonon line\\
     \hline
     H (related) & no & no \\
    \hline
    Li$_i$ & shallow donor & -\\
    Li$_s$ & acceptor & -\\
    \hline
    Na$_i$ & shallow donor & - \\
    Na$_s$ & acceptor & -\\
    \hline 
    \end{tabular}\\
    \footnotesize \noindent X$_s$: X in substitutional site, X$_i$: X in interstitial site
    \label{table::group_1}
\end{table}

\subsubsection{H}

Hydrogen is the most abundant impurity in diamond after nitrogen.
It is commonly found in the chemical vapor deposition (CVD) process when synthesizing diamond, which occasionally leads to its incorporation within the diamond lattice.
Although hydrogen has mainly been associated with absorption peaks in diamond~\cite{h-diamond-fritsch}, luminescence has also been experimentally observed at $512.8$ and $545.8$~\cite{GIPPIUS1983}, $555.0$~\cite{Zaitsev2001_handbook}, $3202.0$~\cite{Fuchs1995_hydrogen}, and $1358.0$, $1370.0$ and $1382.0$~nm~\cite{FUCHS1995_hydrogen-ir}.
Hydrogen vacancy (H$V$) complexes involving nitrogen~\cite{GloverClaire2003Hiid} and oxygen~\cite{KomarovskikhAndrey2014Esot} have been reported.
In general, hydrogen tends to form complexes with other impurity species, which are described in their respective sections.

Calculations of the geometry, defect formation energies and properties of various hydrogen complexes H$_{x}V$, where $x \in [1,4]$, have been performed~\cite{hvacCzelej}, showing that defects enabling C-H bonds are favourable and thus hydrogen can be expected to aggregate with vacancies.
It was further predicted that higher numbers of hydrogen yield a more stable defect.
However, the H$_{4}V$ center is electrically inactive, and thus does not induce a defect state in the band gap.
H$_{1}V$ is predicted to exhibit no luminescence, the H$_{2}V$ center is also neither optically nor EPR active.
The calculations indicate that H$_{3}V$ has a paramagnetic doublet state, therefore being EPR active, but without luminescence.

Further studies indicate~\cite{SHAWM.J2005Ioqt} that quantum tunneling is an important factor in the observed electron paramagnetic resonance (EPR) spectra associated with the nitrogen-hydrogen vacancy complex. 

\subsubsection{Li}

Lithium is a potential shallow donor in diamond, although its behaviour is highly dependent upon the exact site in the lattice. 
Interstitial Li in pristine diamond favors the tetrahedral $T_d$ symmetry with the neighboring C atoms pushed back from the Li~\cite{Goss2007,Lombardi2007a,Lombardi2007b}. 
The interstitial Li is predicted~\cite{Lombardi2007a} to induce very shallow donor states, merging with the diamond conduction band indicating metallic behaviour. 
Close to the surface, the site is energetically more favorable, and thus more stable. 
This behavior has been used to explain~\cite{Kajihara1991} the low solubility of Li in diamond and its clustering at the subsurface region and grain boundaries.  

Substitutional Li shows the lowest energy with a $C_s$ configuration, although the difference is small to $T_d$, which has also been previously reported to have the lowest energy~\cite{Lombardi2007b}. 
The calculations predict~\cite{Czelej2017} that substitutional Li is an acceptor and stable with a charge state of 1+, 0, 1- and 2-. 
Comparatively, the formation energy of substitutional Li is about $1.74$~eV lower than that of the interstitial Li, making it more accessible when vacancies are present. 

When subjected to elevated temperatures, Li can diffuse and be trapped in vacancies, where it will be passivated and can compensate for the remaining Li interstitials, resulting in high resistivity in the sample. 
Nevertheless, without access to vacancies, $n$-type doping can be achieved~\cite{Lombardi2007b}.  
Further computational work predicts~\cite{ZHANG2023,DELUN2020} that co-doping with N or B can improve the solubility of Li into diamond, and some related complexes also show promising behaviour as shallow donors.

\subsubsection{Na}

Sodium is another potential shallow donor in diamond, with properties similar to those of lithium.  
Interstitial Na in diamond is predicted~\cite{Lombardi2007a,Lombardi2007b,Goss2007} to have the lowest energy in a tetrahedral $T_d$ symmetry, with nearest C relaxing away from the dopant.
This configuration induces a shallow donor level within the band gap, deeper than those of Li. 
Similarly as for Li, the Na donor level merges with the conduction band of diamond and the Fermi energy is close to the bottom of the conduction band.
Unlike Li, the interstitial Na has a relatively high activation energy for diffusion.

Substitutional Na in diamond also favors~\cite{Lombardi2007a,Lombardi2007b} the tetrahedral $T_d$ symmetry, although $C_{2v}$ is the lowest in energy near the crystal surface.
Similarly as for Li, substitutional Na in diamond acts as an acceptor~\cite{Czelej2017}. 
The formation energy~\cite{Lombardi2007a} for substitutional Na is $2.78$~eV lower than that of interstitial Na. 

\subsection{Group 2: alkaline earth metals}

The studied impurity configurations in group $2$ consist of beryllium and magnesium complexes.
Both Be and Mg have been been found optically active single photon emitters and thus promising candidates for quantum information technologies.
They lead to $n$- or $p$-type doping depending on the lattice site.
Control of the configurations has to be achieved in order to realise their full potential.

\begin{table}[h]
    \caption{
    {\bf Group 2.} 
    States introduced in the diamond band gap by group $2$ impurity centers are indicated by the nature of the doping: acceptor ($p$-type) and donor ($n$-type). The optical zero phonon transitions are indicated when available.
    }
    \begin{tabular}{ |c|c|c| }
    \hline
    impurity & gap state & zero phonon line (nm)\\
    \hline
    Be$_i$ & shallow donor &- \\
    Be$_s$ & acceptor &-\\
    Be & - & 256\\
    BeH & acceptor/donor &- \\
    \hline
    Mg$_i$ & shallow donor &- \\
    Mg$_s$ & deep acceptor &-\\
    Mg$V^-$ & - & 557.6\\
    MgH & acceptor/donor &-\\
    \hline 
    \end{tabular}\\
    \footnotesize \noindent X$V$: X plus vacancy, X$_s$: X in substitutional site, X$_i$: X in interstitial site, XH: X with hydrogen
    \label{table::group_2}
\end{table}

\subsubsection{Be}

Beryllium doping can generate optically active centers in diamond.
Experiments~\cite{UEDA20081269} with Be implantation find several cathodoluminescence peaks with the zero phonon line at $256$~nm.
Further spectroscopy~\cite{UEDA2009121} with CVD-grown Be-doped diamond reveal a peak at $260.5$~nm.

Computational investigations~\cite{Yan2009} show that interstitial Be favors the tetrahedral $T_d$ site.
It introduces donor states in the diamond band structure that mix with the diamond conduction band.
The Fermi level is found at the conduction band indicating $n$-type behaviour and metal-type conductivity. 

Substitutional Be is predicted~\cite{Yan2009} to find the optimal configuration in the $T_d$ symmetry site.
For this structure, Be impurity states are created above the diamond valence band and the Fermi level is located at these states.
Substitutional Be is therefore an acceptor in diamond and may compensate for the interstitial donor-type Be.
However, spin-polarized density functional theory calculations have also predicted~\cite{Czelej2017} that the $C_{3v}$ site could be the lowest energy configuration for substitutional Be.
Nevertheless, also in this configuration it is a strong acceptor.
It has a low formation energy indicating that it can be easily incorporated in the lattice.

Be interstitial with hydrogen complexes show~\cite{Yan2009} mostly $p$-type, insulating behavior.
Only two configurations have been found to oppose this and instead introduce donor levels in the bad gap.
In the first one, the H atom is located between the nearest and second nearest C atom of the Be, where the Be takes the $C_{3v}$ symmetry site.
In the second one, the H is located between the Be (at the $C_{3v}$ symmetry site) and the nearest neighbor C. 

\subsubsection{Mg}

Magnesium vacancy center (MgV) is optically active and has the potential to act as a bright quantum bit. 

An intense zero phonon line is measured~\cite{Corte2022, LühmannTobias2018Saeo} at $557.6$~nm in ion implanted samples and correlates well with the computationally predicted~\cite{Pershin2021} value for negatively charged MgV$^{-1}$, which was also found to be energetically the most stable of the studied defects. 
Photoluminescence peaks of Mg$V$ have so far been  measured ~\cite{Corte2022} at $574.7$~nm, $599.3$~nm and $643.8$~nm.
The study also indicates an appealing excited-state radiative lifetime of $0.3$~ns. Further doping of the diamond with donors~\cite{Lhmann2019} increases the center's optical emission.

Computational work~\cite{Pershin2021} predicts interesting spin-dependent properties for the MgV ground state configuration. Two doublet and quartet states were identified separated by about $22$~meV, indicating accessible spin-conversions at room temperature.

Mg is a potential $n$-type dopant in diamond.
Interstitial Mg is predicted~\cite{Yan2009,Czelej2017} to take the tetrahedral $T_d$ site in the perfect diamond lattice.
The structure shows similar behaviour with the Be interstitial.
The dopant introduces impurity states near and overlapping with the diamond conduction band.
The Fermi level is found at the conduction band, making the intersitital Mg a $n$-type conductor.  

Substitutional Mg has the lowest energy configuration in the $T_d$ symmetry site.
It introduces~\cite{Yan2009} impurity states above the diamond valence band with the Fermi level located at these sites.
Compared to Be, the impurity states are deeper in the band gap, and show $p$-type acceptor behaviour.
The formation energy is about double that of Be~\cite{Czelej2017}. 
The behaviour of interstitial Mg and hydrogen complexes in diamond are predicted~\cite{Yan2009} to coincide with the behaviour of BeH.

\subsubsection{Ca}
Calcium impurity in diamond has not been extensively researched so far. It can be introduced into diamond using ion implantation ~\cite{LühmannTobias2018Saeo}, and is linked to a weak fluorescence with a zero phonon line at $557.2$~nm.

\subsection{Group 3-11: transition metals}

Transition metals form the largest group of elements studied in diamond.
Many of them have promising properties as optically active single photon emitters, including titanium-, chromium- and nickel-related centers.
Moreover, chromium in diamond is a potential center for spintronic applications and the nickel split-vacancy has been proposed as a solid state quantum bit. 

\begin{table}[h!]
    \caption{
    {\bf Group 3-11.} 
    States introduced in the diamond band gap by the group $3-11$ impurity centers are indicated by the nature of the doping: acceptor ($p$-type) and donor ($n$-type). The optical zero phonon transitions are indicated when available.
    }
    \begin{tabular}{ |c|c|c| }
    \hline
    impurity & gap state & zero phonon line (nm)\\
    \hline
    Ti$V$ & - & 541.4, 760.6\\
    TiN & - & 427.5, 440.3 \\
    Ti$V{^-}$N & - & 502, 503.4, 510.7, 1041.9\\
    Ti (related) & - & 494.5, 970.5, 991.8 \\
    \hline
    V$_s$ & donor & -\\
    \hline
    Cr$V$ & acceptor & -\\
    Cr (related) & - & 756\\
    \hline
    Mn$V$ & deep acceptor & -\\
    Mn (related) & - &  621.2\\
    \hline
    Fe (related) & - & -\\
    \hline
    Co (related) & - & 478.7, 523.8, 544.5, 561.8 \\
    & & 580.7, 623.3, 669.3 \\
    CoSiN & - & 543.1, 554.0, 580.0, 841.1 \\ 
    \hline
    Ni$^{-}_{s}$ & acceptor & - \\
    Ni$V$ & acceptor & 652.5, 1033 \\
    NiN & - & 450, 450.7, 468.1, 473.5, \\
    & & 477.1, 478.4, 478.8, 494, \\
    & & 499.6, 502.2, 510.7, 514.8,\\
    & & 515.7, 518, 522, 527.4, \\
    & & 535.2, 537.3, 553.1, 563.5, \\
    & & 598.7, 622.7, 639.1, 642.6, \\
    & & 658.3, 689.9, 693.7, 703.6, \\
    & & 711, 720.6, 722.6, 752.3, \\
    & & 746.7, 877.45, 832.4, 913.9, \\
    & & 890, 546.6, 552.9, 883.2, 885.6 \\
    NiB & - & 690.5\\
    Ni (related) & - & 403.0, 404.4, 484.3, 491.3, \\
    & & 496, 500.5, 516.2, 539.4, \\
    & & 540.8, 712.6, 727, 732.1,\\
    & & 885.6, 856, 868, $999.9-1024.7$\\
    NiSiN & - & 656.5, 668.6, 677.6, 681.4\\
    \hline
    Cu (related) & - & 709.3\\
    \hline
    Zn (related) & - & 518\\
    \hline
    Mo (related) & - & -\\
    \hline
    Ag (related) & - & 398.5, 401, 405\\
    \hline
    Ta (related) & - & 699, 709\\
    \hline
    W (related) & - & 713\\
    \hline 
    \end{tabular}\\
    \footnotesize \noindent X$V$: X plus vacancy, X$_s$: X in substitutional site, XY: complex
    \label{table::group_3}
\end{table}

\subsubsection{Ti}

Titanium centers have promising optical properties.
They are linked to several paramagnetic resonance centers and can act as single photon emitters. 

Ti impurities have been introduced to natural diamond via ion implantation~\cite{GIPPIUS1993_defect}.
They introduce a zero phonon line at $991.8$~nm.
Two more zero phonon lines at $494.5$ and $970.5$~nm appear after high temperature annealing.
Later on, Ti impurity centers were linked to the paramagnetic resonance centers OK1~\cite{klingsporn1970} and N3~\cite{shcherbakova1972} found already in the 1970s.
Ti split-vacancy with a neighboring N (Ti$V$N) was proposed to give rise to the OK1 center, which has zero phonon lines at $503.4$~nm and $510.7$~nm.
The N3 center with a zero phonon line at $440.3$~nm on the other hand was assigned~\cite{nadolinny2009_new_data} to a Ti substitution with a neighboring N atom, \textit{i.e.}, a substitution dimer.
A band linked to a Ti- and In-containing center has been reported~\cite{Butuzov1976} at $660$ and $720$~nm. 

Computational work on the Ti centers~\cite{Czelej-Ti} shows that the Ti-vacancy complexes have lower formation energies than substitutional configurations.
The accompanying N is found to further stabilise the center. 
The neutral Ti vacancy center is predicted to form a split-vacancy configuration.
In the ground state, the site has a $D_{3d}$ symmetry, but $C_s$ symmetry was also found stable.
The calculated symmetry and electronic structure indicate that optical excitation could possibly lead to ionization, instead of photon emission, which would lower the efficiency of the Ti$V$ center as a single photon emitter.

The Ti$V$N (OK1) is a promising candidate for a single photon emitter.
Interestingly, the lowest calculated excitation energy~\cite{Czelej-Ti} corresponds to red light with the zero phonon line at $1041.9$~nm near the IR region.
This transition has not yet been detected and requires experimental investigation.
The numbers of OK1 centers has been shown to increase in a diamond rich with N3 (TiN) centers after annealing at high temperatures.
This suggests transformation from N3 to OK1.
The TiN substitution dimer is predicted to relax into the $C_{3v}$ ground state symmetry, with the $C_s$ symmetry close in energy.
The calculated zero phonon line~\cite{Czelej-Ti} is at $427.5$~nm and coincides relatively well the experimentally observed one.

\subsubsection{V}

Vanadium-doped diamond is predicted~\cite{Zhang2012} to have several impurity donor levels in the diamond band gap between the conduction band and the gap middle.
However, the structure is associated with a strong distortion of the lattice.
When successfully incorporated, the V substitution should act as a $n$-type dopant.
Further experiments~\cite{Zhang2012} have confirmed the incorporation of the dopant, albeit with a suppressed doping concentration.

\subsubsection{Cr}

Chromium impurities in diamond show promising features for applications in spintronics and optoelectronics. 
Experiments have demonstrated~\cite{Aharonovich2009,Aharonovich2010-b} Cr-based single photon emitters in synthetic diamond at room temperature.
The authors report on an emitter with a zero phonon line at $756$~nm and full width at half maximum of $11$~nm.
Multiple lines between $730-780$~nm are reported in Ref.~\cite{Zaitsev2001_handbook} (and references therein).
The excited state lifetime is short (in the ns-range) and the emission rate is high indicating promise for applications. 

Optically active Cr single photon emitters have been created~\cite{GIPPIUS1983,Aharonovich_2011} via ion implantation, with emission at $741$~nm.
Co-implantation with oxygen increases the number of optical centers, but nitrogen has an essential role in producing the optically active Cr center.
Without N, no single photon emission was found in ultra-pure diamond implanted only with Cr and O.
Calculations show~\cite{Liu2013} that the Cr with a neighboring vacancy with N at the nearest neighbor site introduces strongly spin-polarized impurity levels into the diamond band gap.
These centers are likely to account for the bright single photon emission.
Oxygen incorporation is predicted to broaden the emission line and thus weaken the quality of the luminescence center.

The properties of Cr in diamond have been shown~\cite{Benecha2011} to also strongly depend on the dopant's charge state.
Various studied Cr impurity structures introduced deep donor and acceptor levels in the diamond band gap. 
In $p$-type diamond the most stable configuration was found to be the doubly charged Cr$^{+2}$ with a neighboring vacancy.
The same structure, although with a negative charge state (Cr$^{-2}$), was clearly the favored site in $n$-type diamond. 

\subsubsection{Mn}

Manganese is an optically active center in diamond introducing a deep acceptor level in the band gap.
Mn impurities are likely to couple with vacancies.
The Mn vacancy (Mn$V$) complex as well as the subsitutional site have lower formation energies at the ground state compared to interstitials~\cite{ASSALI2009}, the Mn$V$ being energetically favored.
Magnetically the Mn$V$ is predicted to have nearly degenerate spin configurations and to act as a deep acceptor~\cite{Steven2003}.
Recently~\cite{Kunuku2019}, Mn related color centers have been measured in ion implanted and annealed ultrananocrystalline diamond films.
The zero phonon line emission was detected at $621.2$~nm and characteristic phonon sidebands at $611.2$ and $630.3$~nm.

\subsubsection{Fe}

So far iron has not attracted much attention in diamond research. 
Substitutional Fe in diamond is nonmagnetic~\cite{Chanier2012} and relaxes into the $D_{3d}$ symmetry with a spin of $2$.
With a neighboring vacancy~\cite{ASSALI2009} it has a $T_d$ symmetry with spin $0$.
Compared to these, the interstitial site is the least favored with a significantly higher formation energy.
In this case the impurity assumes a site with $T_d$ symmetry with a spin of $1$.

\subsubsection{Co}

Cobalt impurities exhibit several paramagnetic centers in diamond~\cite{Twitchen2000,Nadolinny2017}.
Photoluminescence measurements reveale~\cite{Lawson1996} vibronic systems with zero phonon lines located at $478.7$, $523.8$, $544.5$, $561.8$, $580.7$, $623.3$ and $669.3$~nm.
Annealing revealed~\cite{JOHNSTON2000} that many of the lines appear at the temperature regime of nitrogen aggregation which lead to the conclusion that the optically active centers are likely to be related to CoN complexes.
The proposed models~\cite{JOHNSTON1999647} include a substitutional Co and a Co-vacancy complex with a N at the nearest neighbor site.
In addition, two other Co-related lines (at $560.9$ and $842$~nm) are reported in Ref.~\cite{Zaitsev2001_handbook}.
In the presence of N and Si, the Co-related center has been linked~\cite{SITTAS1996} to several zero phonon lines at $543.1$, $554.0$, $580.0$ and $841.1$~nm.  

Calculations show~\cite{JOHNSTON2002} that the energetically most favorable configuration is the Co-vacancy complex, while interstitial and substitutional configurations are less likely due to higher long-range elastic strain.
Substitutional Co has $T_d$ or $D_{2d}$ symmetry, depending on the charge state.
The spin varies from $0$ to $1/2$, $1$ and $3/2$, likewise for Co in the interstitial site~\cite{Larico_2008}.

\subsubsection{Ni}

Nickel can be easily incorporated into synthetic diamond and it can also work as a catalyst during the growth.
The formation of Ni centers and their optical properties are reviewed in Refs.~\cite{Nadolinny2017,COLLINS2000}.
The Ni impurities in diamond have promising photoluminescence properties and multiple gap states.
Luminescence lines that appear in both implanted and synthetic diamond are observed~\cite{GIPPIUS1983} at $484$, $451$ and $885$~nm. 
Further studies show multiple zero phonon lines, including $403.0$ and $404.4$~\cite{Zaitsev2001_handbook}, $484.3$~\cite{Collins_1983}, $491.3$~\cite{Lawson1993_annealing}, $496$~\cite{Zaitsev2001_handbook}, $500.5$~\cite{YELISSEYEV1995_photo}, $516.2$ and $539.4$~\cite{KUPRIYANOV19991_photo}, $540.8$~\cite{zaitsev2000vibro}, $712.6$~\cite{orwa2010_Ni}, $727$~\cite{Lawson1996}, $732.1$~\cite{Lawson1993_annealing,YELISSEYEV1995_photo}, $885.6$~\cite{GIPPIUS1983}, $856$ and $868$~\cite{Zaitsev2001_handbook} and $1000-1025$~nm~\cite{Lawson1993_new_nickel}.  
High-temperature annealing above $1700^\circ$C leads to multiple new absorption lines~\cite{NADOLINNY1994_new_param,Lawson1993_annealing}.
While measurements of the electron paramagnetic resonance in Ni-doped diamond took place early on~\cite{Samoilovich1971,Lounser1966}, the exact assignment of each  center and the corresponding Ni configuration (mainly consisting of the interstitial, the Ni vacancy complex (Ni$V$) and the substitutional Ni as well as their varying charge states) has been ongoing ever since. 

Based on simulations, Ni$V$ relaxes into the split-vacancy configuration. 
It has the potential to be applied as a solid state quantum bit, and has recently been reported~\cite{Bourgeois2017} to exhibit two bands near infrared, at about $652.6$ and $1033.2$~nm.
These states were first discovered during photocurrent measurements~\cite{Londero2018} with the N$V$ center, and were later assigned to rise from the Ni$V$ center and its two acceptor levels.
Computational magneto-optical spectroscopy~\cite{Thiering2021} has been used to further shed light on the Ni$V$ center.
The negative charge state of Ni$V$ was connected~\cite{Isoya1990} to the $885.6$~nm optical center and the so-called NIRIM-$2$ electron paramagnetic resonance center, which has been studied in detail and its origin keenly discussed~\cite{JMBaker_2003,Isoya1990,Mason1990}.
According to calculations, the Ni$V^-$ has spin $1/2$ and a long spin coherence time at cryogenic temperatures.
The spin can be optically initialised and read out, marking it as a serious competitor to the well known N$V$ center.

Beyond the Ni$V$ center, Ni-doped diamond has multiple optical absorption bands and it exhibits distinct zero phonon lines. 
Computational studies reveal~\cite{Chanier2013} the negatively charged substitutional Ni in diamond to have a spin state of $3/2$, with an acceptor gap state at about $2$~eV above the valence band maximum.
This impurity center has been assigned to the experimentally detected W8 electron paramagnetic center. 
Also other Ni centers have been detected by electron paramagnetic resonance, but the correct identification of any of these centers and their respective paramagnetic properties is not straightforward~\cite{Thiering2021,Larico2004,Larico2009}. 
The interstitial Ni site (spin $1/2$) has been calculated to have a very high formation energy~\cite{Larico2009} coupled to a low diffusion barrier (below $1$~eV)~\cite{JPGoss_2004}, indicating that the defect is highly mobile at elevated temperatures. 
The $885.6$~nm ($1.40$~eV) center has been linked to the interstitial N$^+$ atom by uniaxial stress and Zeeman measurements~\cite{nazare1991_optical}.
Nevertheless, none of these centers possess similar bright photoluminescence as that evident for the NiV center, and required for a qubit.
Near surface, \textit{e.g.}, in size-controlled nanodiamond structures, the Ni-related valence band can be perturbed leading to a strong quantum confinement effect.
Such systems have been proposed for \textit{in vivo} multicolor bioimaging~\cite{Thiering2014}. 

In addition to pure Ni defects, also several co-dopant configurations exist.
One of them is the NE8 complex that consists of one Ni atom surrounded by four N atoms.
NE8 can generate single photons emitted at $1033.2$~nm~\cite{Gaebel_2004}, and can be excited optically at room temperature.
The spectrally narrow and intense emission makes this center well suited for generating single wavepackets for optical quantum computing.
The zero phonon line is $885.6$~nm~\cite{helena1991}, near infrared, indicating feasibility towards quantum cryptography.
In confined structures, such as nanodiamonds, even higher single-photon detection efficiency has been reported~\cite{Wu_2007}.
Other Ni complexes include pairs with B and N~\cite{Larico2009} in combination of interstitial and substitutional pairs.
NiN complexes are commonly formed during high temperature annealing when N, that can occur naturally in diamond, becomes mobile and pairs with the less mobile Ni atoms.
The N incorporation may substantially alter the electronic structure of the complex, and not simply donate electrons.
Several absorption lines are linked to Ni and N containing impurity centers including the zero phonon lines at $546.6$, $552.9$ and $883.2$~nm~\cite{YELISSEYEV1995_photo}, as well as at $450$, $450.7$, $468.1$, $473.5$, $477.1$, $478.4$, $478.8$ $494$, $499.6$, $502.2$, $510.7$, $514.8$, $515.7$, $518.0$, $522$, $527.4$, $535.2$, $537.3$, $553.1$, $563.5$, $598.7$, $622.7$, $639.1$, $642.6$, $658.3$, $689.9$, $693.7$, $703.6$, $711.0$, $720.6$, $722.6$, $752.3$, $746.7$, $877.45$, $832.4$ and $913.9$~nm~~\cite{Zaitsev2001_handbook,Nadolinny_1999_nickel,YELISSEYEV1996,YELISSEYEV1995_photo,KUPRIYANOV19991_photo}, as well as the broad band at $890$~nm~\cite{collins1990_segregation}.
The $690.5$~nm zero phonon line has been attributed to a B- and Ni- containing center as discussed in Ref.~\cite{Zaitsev2001_handbook} (and references therein).
On the other hand, Ni, N and Si containing centers have been linked~\cite{SITTAS1996} to the $656.5$, $668.6$, $677.7$ and $681.4$~nm zero phonon lines.

\subsubsection{Cu}

Literature on copper-doped diamond is thin.
A computational study~\cite{Chanier2012} reported that Cu in a substitutional site, relaxed to the $T_d$ symmetry, has a magnetic ground state with a total magnetic moment calculated at $3~\mu_\mathrm{B}$ that corresponds to a total spin of $3/2$.
Among transition metals, the magnetic moment is reported to be the highest one.
Experimentally~\cite{KUPRIYANOV2016198}, when Cu was used as a forming agent during diamond synthesis, the photoluminescence spectrum showed a new zero phonon line at $709.3$~nm.
The team tentatively assigned this to Cu impurities, but confirmation is lacking.
In any case, further work is needed to confirm the magneto-optical properties of Cu-doped diamond.

\subsubsection{Zn}

Zinc impurities in diamond have not attracted much attention so far, although implanted and annealed Zn centers have been reported to lead to a zero phonon line at $518$~nm~\cite{GIPPIUS1983,Zaitsev2000}.
The impurities introduce very deep impurity bands in the band gap~\cite{ALVES1992}.

\subsubsection{Mo}

Molybdenum can be introduced into diamond with a high concentration during growth~\cite{biener_2010}. 
Rutherford back scattering and channelling experiments indicate that Mo sits neither on a substitutional nor an interstitial site. 
The authors suggest that more complex metal vacancy clusters are likely formed, where the substantially larger Mo displaces multiple carbon atoms to relieve the compressive strain around itself. 
Photocurrent measurements with Mo-containing diamond show four new peaks at $1087.6$, $1662.0$, $1761.1$ and $2421.6$~nm~\cite{kromka2003_detection}. 
X-ray photoemission spectroscopy confirmed the Mo impurities, although measurements with Rutherford backscattering and electron paramagnetic resonance could not confirm this.   

\subsubsection{Ag}

Silver is an unlikely candidate for diamond based applications, although it introduces an optical center with a zero phonon line at $398.5$~\cite{Zaitsev2000} and additional lines at $401$ and $405$~nm~\cite{GIPPIUS1983}.
The impurities were introduced via ion implantation and subsequent annealing.

\subsubsection{Ta}

Tantalum is optically active in diamond~\cite{Harris1996}.
A sharp zero phonon line is measured~\cite{zaitsev2000vibro} at $699$~nm accompanied by a weak band at $709$~nm.
 
\subsubsection{W}

Tungsten-related impurity centers in diamond show multiple zero phonon lines, of which the most intense one is at $713$~nm~\cite{Zaitsev2000}. 
Computational work has indicated that W substitutions have potential use in spintronic and microelectromechanical systems.
Simulations show~\cite{SU2022} clear magnetic behavior.
This site is energetically favored over the interstitial one, as would be expected of such a large atom.
The calculated band structure of the spin-up and -down configurations show two different aspects.
With spin-up the band structure is metallic, while with spin-down a band gap appears.
Such half-metallic properties coupled with stable ferromagnetism could give rise to applications. 
Interestingly, W can achieve high doping levels in diamond~\cite{biener_2010}.
Recently~\cite{braun2023}, highly W-doped diamond was used to reach a record neutron yield in inertial confinement fusion experiments.
The process critically depends on the properties of the ablator shell, which is used in the target of the reaction.
In the diamond shell, the W impurities selectively filter non-thermal hard rays that would otherwise harm the process.

\subsection{Group 13}

Of group $13$ impurity centers, boron is the most promising one.
The prominent acceptor is easily doped into diamond with high concentrations.
Applications vary from electrochemistry to microelectrodes.
The B-vacancy complex is single photon emitter and has been proposed as a solid state quantum bit.

\begin{table}[h!]
    \caption{
    {\bf Group 13.} 
    States introduced in the diamond band gap by the group $13$ impurity centers are indicated by the nature of the doping: acceptor ($p$-type) and donor ($n$-type). The optical zero phonon transitions are indicated when available.
}
    \begin{tabular}{ |c|c|c| }
    \hline
    impurity & gap state & zero phonon line (nm)\\
    \hline
    B (related) & shallow acceptor & 534.4, 472, 515.8, \\
    & & 777.8, 517.5, 259.5\\
    \hline
    Al$_s$ & deep acceptor & - \\
    \hline
    Ga$V^-$ & - & 679\\
    \hline
    In$V^-$ & - & 584 \\
    \hline
    Tl (related) & - & 614\\
    Tl$V^-$ & - & 437\\
    \hline 
    \end{tabular}\\
    \footnotesize \noindent X$V$: X plus vacancy, X$_s$: X in substitutional site
    \label{table::group_13}
\end{table}

\subsubsection{B}

Early on, boron was determined~\cite{poferl-1973} to be the dominant $p$-type acceptor in laboratory-grown diamond that can selectively filter photons with energies higher than $1.8$~keV.
It supports very high doping levels with densities reaching $8 \times 10^{20}$~cm$^{-3}$~\cite{LAGRANGE19981390} (compared to the intrinsic density of diamond of $2 \times 10^{23}$ C atoms per cm$^3$).
Heavily doped samples show nearly metallic behavior.
This has lead to B-doped diamond being widely studied in electrochemistry, especially for its use in microelectrodes~\cite{Cobb2018,Macpherson2015,Muzyka2019}.
Currently, B-doped diamond is one of the leading materials for electrochemical generation of strong oxidising radicals that destroy organic pollutants in aqueous solutions.
It is used for example in disinfection and wastewater treatment~\cite{Sires2014}. 
B doping changes radically the optical properties of diamond at the visible range. 
Electrodes made of heavily B-doped diamond are now also applied for electrochemical X-ray fluorescence~\cite{Hutton2014} and electrochemical ultraviolet-visible-infrared spectroscopy~\cite{Yingrui2007}. 
Moreover, heavily B-doped diamond exhibits superconductivity~\cite{Ekimov2004,Takano2004}.

In optical measurements, the zero phonon line of B-doped diamond appears at $534.4$~nm~\cite{Ruan1992} and in some cases also at $472$, $515.8$ and $777.8$~nm~\cite{Zaitsev2001_handbook} and at $517.5$~nm~\cite{Freitas1993}, and at $259.5$~\cite{Walker1979}.
The so-called Green Band at $540$~nm~\cite{GHEERAERT1994} has also been detected in B-doped polycrystalline diamond.
Optical properties of B-doped diamond are extensively discussed in Ref.~\cite{Zaitsev2001_handbook}. 

The B-vacancy (B$V$) complex has been identified as a potential quantum bit.
The complex has the $C_{3v}$ symmetry, where the B atom stays in the substitutional site.
Among different charge states, the negatively charged B$V^{-}$ has risen to be the most promising candidate for quantum applications.
Interestingly, for B$V^{-}$, the up and down spin states are separated on either side of the Fermi level.
This makes the center a suitable candidate for a quantum bit as well as for single photon emitters.
The B$V^{-}$ center introduces multiple deep gap states in the diamond gap found at $2.820$, $-0.722$, $-0.722$ and $1.939$~eV for the spin-up, and at $2.571$, $-0.722$, $-0.722$ and $1.983$~eV for the spin-down states~\cite{MURUGANATHAN2021108341}.
Such gap states do not occur for the substitutional B dopant. 

In a substitutional site, neutral B relaxes into the $C_{3v}$ symmetry, while in a negative charge state B$^-$ it prefers the $T_d$ structure~\cite{Goss2006}, where all the electrons can have full covalent nature.
Substitutional B introduces shallow acceptor levels in the bad gap with an ionization energy of about $0.38$~eV~\cite{Bourgoin1979}.
The recently calculated acceptor level~\cite{Czelej2017} is found at $0.39$~eV above the valence band maximum.

Aggregation of substitutional B is unlikely and requires very high temperature due to high calculated migration barriers.
Any B pairs would induce deeper acceptor levels as compared to isolated substitutions, and thus would make a weaker contribution to the increased conduction of B-doped diamond.
On the other hand, agglomerations of multiple B with a vacancy can have a positive effect in the metallic conductivity of the heavily B-doped diamond~\cite{Goss2006}, although due to high formation energies such structures are unlikely to form under equilibrium conditions.

\subsubsection{Al}

The aluminum impurity has not yet been found to show any remarkable properties in diamond.
As a substitutional impurity~\cite{Goss2005,Czelej2017}, Al has energetically the most favored configuration in the tetrahedral $T_d$ symmetry.
It induces partially filled impurity states at the top of the diamond valence band and acts as a deep acceptor~\cite{LOMBARDI20081349}. 
The calculated charge density of substitutional Al in diamond shows~\cite{Barnard2003} the presence of ionic-like bonding between the dopant and carbon atoms.
No Al-C bonds were detected in the charge density map, but close to the location of the missing bonds an increase of charge is visible, suggesting that the Al has donated electrons. 
In a split-vacancy configuration, Al is predicted to be optically active~\cite{Goss2005}.
In the substitutional site with a neighboring vacancy, Al takes the $D_{3d}$ symmetry site. 
Experiments with CVD-grown heavily Al-doped diamond~\cite{Mori2015} show low activation energy at room temperature.
Luminescence has been detected at $415$ and $490$~nm in synthetic diamond with Al dopants, as cited in Ref.~\cite{Zaitsev2001_handbook}.

\subsubsection{Ga}

Gallium impurities have also not shown great promise yet, but it is worth noting that Ga ions are regularly used in focused ion beams for milling and modifying the diamond surface, and may thus be present in processed samples.
Computational investigation~\cite{Goss2005} reveals that Ga relaxes to a tetrahedral $T_d$ symmetry in diamond.
It opens new gap states at $1.4$~eV above the diamond valence band maximum.
The dopant exerts a compressive strain on the surrounding lattice.
In a vacancy complex, Ga$V$ assumes the split-vacancy site with a $D_{3d}$ symmetry.
Due to the inversion symmetry of the site, the complex can not have a net dipole moment and its optical transitions are not sensitive to charge noise.
The zero phonon line has been calculated~\cite{Harris2020} for the negatively charged  Ga$V^-$ at $679$~nm.
Experimental studies show~\cite{DRAGANSKI201347} that when implanted, Ga changes the refractive index of diamond.   

\subsubsection{In}

According to computational work~\cite{Doyle1998}, In embedded in diamond prefers a site with a neighboring vacancy and relaxes into a split-vacancy structure.
Calculations~\cite{Harris2020} with the negatively charged split-vacancy In$V^-$ indicate a zero phonon line at $584$~nm.
Substitutional site has the $T_d$ symmetry and gap states $1.8$~eV above the valence band maximum~\cite{Goss2005}, but is energetically less favorable with a positive binding energy indicating an unstable configuration.

\subsubsection{Tl}

Thallium-doped diamond has been produced via ion implantation and subsequent annealing~\cite{GIPPIUS1983,Zaitsev2000}.
The resulting dopant configurations show strong cathodoluminescence with a zero phonon line of $614$~nm. 
Vibronic analysis indicates that the dopant occupies a tetrahedral position. 
The dopant does not cause large changes in the surrounding lattice nor in the density distribution of short wavelength optical phonons~\cite{Zaitsev2000}. 
Another configuration predicted by computational work~\cite{Harris2020} indicates a stable split-vacancy site. 
In the negative charge state this complex is predicted to have a zero phonon line at $437$~nm.

\subsection{Group 14}

Group $14$ elements include multiple highly promising quantum emitters, such as the tin and lead centers, and also the extensively studied silicon-vacancy center.
The Si$V$ center is an emerging competitor to the well known N$V$ center, with properties that in some aspects surpass those of the N$V$ center.
Germanium centers have also been shown to exhibit similar photoluminescence, making this system another possible candidate for diamond optoelectronic and quantum applications.

\begin{table}[h!]
    \caption{
    {\bf Group 14} 
    Optical activity in group $14$ impurity centers.}
    \begin{tabular}{ |c|c|c| }
    \hline
    impurity & zero phonon line (nm) & luminescence (nm) \\
    \hline
    Si$V$ & 946 & - \\
    Si$V^-$ & 737 & - \\
    Si (related) & - & 733, 738.2, 756.5, \\
    && 756.3\\
    \hline
    Ge$V$ & 602 & -\\
    \hline
    Sn$V$ & 619 & 593.5, 620.3, 630.7, \\
    && 646.7 \\
    \hline
    Pb$V$ & 517 & -\\
    Pb (related) & 520 & 539.4, 552.1, 556.8,  \\
    && 565-600\\
    \hline 
    \end{tabular}\\
    \footnotesize \noindent X$V$: X plus vacancy
    \label{table::group_14}
\end{table}

\subsubsection{Si}

Silicon-vacancy (Si$V$) impurities are one of the most studied point defects in diamond, motivated by the search for candidates for quantum bit realization without the drawbacks that are present in the NV center. 
Sharing many similarities with this defect, silicon-vacancy impurities exhibit multiple favorable properties for various quantum sensing and computing applications. 
Calculations predict that this center relaxes into the split-vacancy structure with $D_{3d}$ symmetry~\cite{GaliAdam2013Aeis,Rogers2014} oriented along the [111] axis. 
Two charge states for this impurity have been intensively studied, namely the (singly) negative and neutral state.

The electronic structure of the negatively charged center (Si$V^-$), has been predicted to be a $S=1/2$ spin system, with a limited spin coherence time of $35-45$~ns~\cite{BeckerJonasNils2017CPaQ}.
However, it exhibits a sharp zero phonon line at $737$~nm with a width of $0.7$~nm and single photon emission at room temperature~\cite{NeuElke2011Spef}, with a high stability of the optical transitions with regards to external electric fields and noise~\cite{Sipahigil2014}.
It possesses superior fluorescent qualities with $70\%$ of its fluorescence concentrated in the zero phonon line~\cite{NeuElke2012Etos}.
Microwave-based control is possible with optically detected magnetic resonance.

The neutral Si$V$ center is photoluminescent~\cite{GreenBL2019Esot} and exhibits properties combining desirable aspects of the nitrogen vacancy and negatively charged silicon vacancy, such as the high Debye-Waller factor, where the majority (up to $90\%$ for $T=4$~K) of the emitted photons fall within the primary zero phonon line at $946$~nm even at higher temperatures.
It was shown to be a paramagnetic spin $S=1$ system. 
Optically detected magnetic resonance has been reported for this impurity complex, with a spin coherence time of $55.5~\mu$s~\cite{ZhangZi-Huai2020ODMR}. 
Although these properties are promising, there is one disadvantage: The charge state is not stable in high purity diamonds. 
A way to mitigate this is to use hydrogen terminated diamond surfaces, which stabilize the neutral charge state of the impurity~\cite{ZhangZi-Huai2023NSVC}. 
The formation energy is $6.9$~eV~\cite{GaliAdam2013Aeis}.

Other configurations such as substitutional and interstitial silicon have been studied theoretically~\cite{JPSi}.
However, they are predicted to be optically inactive, with the latter configuration even being unstable.
The study also reported on silicon hydrogen structures such as the neutral Si$V$H complex, identifying this as the microscopic structure behind the KUL3 signal reported in electron spin resonance studies~\cite{cvdKUL} of impurities associated with silicon and hydrogen.
Calculations on various SiVH$_{x}$ complexes\cite{sivhcompl} (where $x \in [1,5]$), predict that the negatively charged Si$V$H has properties favorable for quantum bit applications, with a formation energy of $4.82$~eV, creating a deep level in the band gap.
The formation energy is lowered even further for complexes with higher numbers of hydrogen.
The neutral SiVH was predicted to have a zero phonon line at $1305$~nm.
Beyond the luminescence of the SiV centers, the $733.0$, $738.2$, $753.5$ and $756.3$~nm zero phonon lines have been linked~\cite{KIFLAWI1997,SITTAS1996,BILODEAU1993} to the Si impurity.

Diamond containing Si$V$ centers can be fabricated by utilizing doping during CVD growth or ion implantation.
In both cases, annealing is necessary to facilitate recombination of thermally activated mobile vacancies with silicon impurities, forming the Si$V$ centers.
It has been reported~\cite{LagomarsinoStefano2018Opos} that for the fabrication of single photon emitter Si$V$ centers low implantation fluences are necessary (in the range of $10^8-10^9$~cm$^{-2}$).
Si$V$ centers were found to show almost identical optical properties independent of the implantation depths after annealing to $1000-1150^\circ$C.
The study also reported that the thermal activation of the Si$V$ centers was enhanced by defects in the diamond matrix.
The incorporation of the Si$V$ defects into the diamond lattice during CVD growth~\cite{LobaevMikhailA2021Iosc} is, however, the preferential fabrication method, because there is little to no damage to into lattice.
It is also worth noting that most CVD-grown diamonds will inevitably contain silicon, as it is often present in the materials of the growth chamber.

\subsubsection{Ge}

Germanium-vacancy complexes (Ge$V$) have been experimentally determined~\cite{Iwasaki2015} to be single photon sources, exhibiting a sharp photoluminescence band with a zero phonon line at $602$~nm, which splits into two components at a temperature of $10$~K.
The excited state lifetime has been estimated to be $1.4-5.5$~ns.
The luminescence was only observed after annealing to $800^\circ$C.
The peak retains a narrow shape even at room temperature~\cite{PalyanovYuriN2015Ganc}, however the percentage  of the intensity falling within the peak is $34\%$, while the rest occupies the vibronic wing. 
At a temperature of $10$~K this ratio rises to $68\%$~\cite{BoldyrevK.N.2018Blod}.  

Calculations~\cite{NadolinnyV.A.2016Esog} predict that the Ge$V$ center relaxes into the split-vacancy configuration with $D_{3d}$ symmetry.
The center is paramagnetic with a spin of $S=1$, with axial symmetry along the [111] axis.
For negatively charged GeV centers, a spin coherence time of $T=19$~ns has been reported~\cite{SiyushevPetr2017Oamc}.
Substitutional germanium (Ge$_{S}$) and germanium with two neighboring vacancies (Ge$V_{2}$) have been studied computationally~\cite{QiuDongchao2021Team}.
The substitutional configuration has been predicted to have little impact on the electronic and magnetic properties of the diamond host, and only in combination with vacancies is a change in the optical and electronic properties expected.
The resulting formation energies for Ge$_{S}$, Ge$V$ and Ge$V_{2}$ are reported to be $7.88$~eV, $7.97$~eV and $10.79$~eV, respectively.

\subsubsection{Sn}

Optically active tin-vacancy (Sn$V$) centers have been created using various methods.
The first studies showed~\cite{Iwasaki2017,Tchernij2017} impurity atom configurations created via ion implantation.
Subsequent high temperature annealing under high pressure was reported to be efficient in healing defects that are unavoidably linked to the implantation process. 
At room temperature, Sn$V$ shows a sharp zero phonon line at $619$~nm, while at cryogenic temperatures this is split into four peaks~\cite{Iwasaki2017}.
In the ground state, the levels show a promising large splitting of about $850$~GHz. 
Photoluminescence spectra of implanted and annealed Sn impurities in diamond~\cite{Tchernij2017} shows emission peaks at $593.5$, $620.3$, $630.7$ and $646.7$~nm.
Sn$V$ has also been introduced into single-crystal diamond during growth under high pressure and high temperature conditions~\cite{PALYANOV2019,EKIMOV2018223}.
The impurity centers are evident in photoluminescence spectra.
The incorporation of Sn centers via growth promises further control over the lattice disorder that is unavoidably linked to the ion implantation process and not completely healed even after annealing.
Lighter elements, such as H and O were reported to prevent Sn doping.

Calculations indicate~\cite{Iwasaki2017} that the Sn$V$ complex takes the split-vacancy configuration with the $D_{3d}$ symmetry. 
Comparison between the ground state energy levels for the neutral and the negatively charged impurities indicate that the Sn$V$ will likely be in a negative charge state.
The Sn$V^-$ can be photoexcited with visible light to the charge state of $2-$~\cite{Thiering2018}, and violet and blue light can ionize the Sn$V^{2-}$ back to Sn$V^{-}$. 

\subsubsection{Pb}

Quantum emission has been achieved with lead-implanted diamond~\cite{Tcherni2018,Trusheim2019}.
Photoluminescence shows~\cite{Tcherni2018} intense emission at $552.1$ and $556.8$~nm that are assigned to the same defect with additional less intense lines at $539.4$ and $574.5$~nm, as well as a band between $565$~nm and $600$~nm.
Of these, the $574.5$~nm line was attributed to the neutral N$V$ center and not to a Pb impurity.

Pb emitters have been successfully isolated by fabrication of nanopillars~\cite{Trusheim2019} using electron beam lithography and reactive ion etching.
The measured zero phonon line was assigned to a doublet at $520$~nm.
As usual, annealing was used to heal defects that were created during the ion implantation.
Interestingly, binary collision approximation-based simulations predict that every Pb projectile produces about $2000$ vacancies when impinging the material at the applied energy ($350$~keV).
Some of these defects can also contribute to the emission.

Computational results suggest~\cite{Thiering2018} that Pb recombines with a vacancy leading to a split-vacancy (Pb$V$) configuration with the $D_{3d}$ symmetry.
While negative charge state is possible given that the diamond contains enough N to donate an electron, Pb$V^-$ can be easily converted into a neutral Pb$V$ by illumination with visible light.
The calculated zero phonon line is located at $517$~nm, confirmed by two independent studies~\cite{Thiering2018,Trusheim2019}.
Overall, the Pb impurity has taken its position among the promising quantum emitters of the group $14$ elements.

\subsection{Group 15: pnictogens}

The reigning status as the most prominent impurity center in diamond is still held by the nitrogen-vacancy pair.
Some of the properties that have lead to the significant interest in this center include its use as a quantum bit, as a single photon source in the visible range and as a detector of external magnetic fields at room temperature.
Other interesting group $15$ impurities, such as phosphorus, arsenic and antimony, indicate the possibility of shallow $n$-type doping that is supported by theoretical calculations.

\begin{table}[h!]
    \caption{
    {\bf Group 15.} 
    States introduced in the diamond band gap by the group $15$ impurity centers are indicated by the nature of the doping: acceptor ($p$-type) and donor ($n$-type). The optical zero phonon transitions are indicated when available.
    }
    \begin{tabular}{ |c|c|c| }
    \hline
    impurity & gap state & zero phonon line (nm)\\
    \hline
    N$V^-$ & - & 637\\
    N$V$ & - & 575 \\
    N$_s$ & acceptor & 271.4, 295.4, 295.8, 296.2, 305.4\\
    NH & donor &-\\
    \hline
    P$_s$ & shallow donor & -\\
    P$V$ & deep acceptor &-\\
    PH & acceptor &-\\
    \hline
    As$_s$ & donor &-\\
    AsH & deep donor &- \\
    \hline
    Sb$_s$ & shallow donor &-\\
    SbH & deep donor & -\\
    \hline 
    \end{tabular}\\
    \footnotesize \noindent X$V$: X plus vacancy, XH: X with hydrogen, X$_s$: X in substitutional site
    \label{table::group_15}
\end{table}

\subsubsection{N}

The nitrogen-vacancy (N$V$) center is a promising candidate for a qubit.
It relaxes into the $C_{3v}$ symmetry oriented along the diamond [$111$] axis.
It is probably the most extensively studied impurity centers in diamond~\cite{Gali_2019,DOHERTY20131}.
The N$V$ center's prominence comes from the complex electronic structure and paramagnetic properties that are related to the electron spin and the nitrogen nuclear spin, their hyperfine interactions and the long electron spin coherence time~\cite{Gruber1997,Bar2013}.

The two states of the single electron spin that realize the quantum bit can be initialized and manipulated via microwave fields and optical excitation and read in a single-shot readout~\cite{Robledo2011}.
The quantum information of the readout signal can be coded in the polarization of the emitted photon or the photon emission at a given frequency.

The N$V$ center can accept an electron becoming N$V^-$.
This can happen via nearby N substitutions that act as donors allowing transitions between charging and discharging~\cite{Larsson2008} or via nearest neighbor co-dopants with sufficient electronegativity~\cite{schwingen2011}.
N$V^-$ demonstrates~\cite{davies1976} a strong optical transition with a zero phonon line at $637$~nm ($1.945$~eV).
The $S=1$ ground state electron configuration and optically active excited states~\cite{Yuchen2010} result in the emission and absorption spectra to be located in the visible range.
The gap states give the N$V^-$ its function as an optically active color center and as a single photon emitter at room temperature~\cite{Kurtsiefer2000} with applications in quantum cryptography~\cite{Beveratos2002}. 
The center can be excited with green light resulting it to emit photons at the wavelength of $700-900$~nm. 

The N$V^-$ center is paramagnetic~\cite{Preez1965}. 
Due to hyperfine interactions~\cite{Gali2008}, N$V$ centers can interact with other paramagnetic impurities in the lattice, such as the $^{13}$C lattice site that can be found in high purity diamond, leading to different polarization properties~\cite{Smeltzer_2011}. 
This makes the N$V$ center very sensitive to the surrounding environment and thus introduces challenges for its use in applications.
Nevertheless, the N$V^-$ center has already been successfully used to detect nanoscale magnetic fields at room temperature~\cite{Balasub2008,Maze2008}. 

In the neutral state (N$V^0$) the N$V$ center has a strong photoluminescence at $575.07$~nm ($2.156$~eV)~\cite{GDavies1979}.
The first electron paramagnetic resonance measurements indicated a S$=3/2$, a metastable excited state~\cite{Felton2008}.
The ground state electron paramagnetic resonance was detected in 2019~\cite{Barson2019}. 
The neutral center provides an intriguing addition to the potential of the nitrogen center~\cite{Gali2009}, such as a magnetometer with sensitivity surpassing the N$V^-$. 
Further, the N$_3V$ impurity complex has been linked to the $N1$ center at $826$~nm as discussed in Ref.~\cite{Zaitsev2001_handbook}.   

Other configurations for nitrogen in diamond include the substitutional site. It introduces acceptor levels in the band gap~\cite{Jones2009_acceptor}. 
The zero phonon lines of a single substitutional N center are measured~\cite{carvalho87_N} at $271.4$, $295.4$, $295.8$, $296.2$, $305.4$~nm. 
However, complexes with H are predicted to have $n$-type behaviour. 
The interstitial site is energetically less favorable than the N$V$ center, but can be achieved via irradiation. This site can introduce infrared and electronic transitions~\cite{Goss2004,MAINWOOD1999,Kiflawi1996}.

Spatial manipulation of the color center has been suggested~\cite{Todd2014} indicating that annealing in the presence of strain can be used to align the N$V$ centers along a crystallographic orientation.
A $89\pm 7 \%$ efficiency was shown in aligning the color centers along the [111]-orientation under $2\%$ compressive biaxial train and annealing at $970^\circ$C.

Further work to understand the complex properties and interactions, as well as control of the orientation, spatial location and surrounding environment is required to realize and improve the performance of quantum devices based on the NV center. 
Specifically, despite its prominence, the N$V$ center has some drawbacks that have spurred the research of other impurity centers in diamond.
One of such features is the low fraction of luminescence located at the zero phonon line (about $4\%$ of the total fluorescence), while most of it is located at the large phonon sideband of the N$V$ center.
The N$V$ center is also susceptible to external noise.
Other impurity centers with large sharp zero phonon emission and resistivity to external noise come from the group $14$ elements.
Nevertheless, the long spin coherence time related to the NV center makes its status prominent and hard to challenge.  

\subsubsection{P}

Recently, phosphorus-doped diamond was found to exhibit remarkably long spin-coherence times at room temperature exceeding even those of the N$V$ center~\cite{Herbschleb2019}.
Phosphorus is a very shallow $n$-type donor in diamond~\cite{Koizumi1997}, introducing gap levels below the diamond conduction band~\cite{Nesladek1999,Barnard2003}.
This makes P-doped diamond a promising material for high-power and high-temperature semiconductor devices.
Experiments have shown that the impurity mainly assumes the substitutional lattice site~\cite{Masataka2001}.
Luminescence lines at $650$~nm~\cite{nijenhuis_p} as well as at $278.6$, $310$, $427.5$, $516.6$, $590.4$~nm~\cite{NAIDOO1999} and $470$~nm~\cite{Zaitsev2001_handbook} have been linked to P doping.
Ion implantation of phosphorus into diamond with subsequent annealing has been reported to result in a low yield of luminescent P-related centers\cite{LühmannTobias2018Saeo}, where two types have been observed: zero phonon line at $557$~nm and doublet lines at $579.4$~nm and $597.2$~nm.

Computational work indicates either $T_d$~\cite{Wang2002} or $C_{3v}$ symmetry~\cite{Goss2005} in the substitutional configuration.
The calculated formation energy is high~\cite{Wang2002} combined with a relatively low binding energy~\cite{Anderson1993} indicating a low solubility in bulk diamond.
Nevertheless, experiments have shown that P can be successfully introduced into diamond from the gas phase during growth~\cite{Koizumi1997,Kato_2007_n-type} with significant concentrations~\cite{Masataka2001}.

Even after P doping, diamond may remain insulating.
It has been argued to result from the formation of P-vacancy (P$V$) complexes, which are deep acceptors and can therefore compensate the $n$-type conductivity.
In the presence of vacancies, it is energetically favorable for the P to recombine with a vacancy occupying the split-vacancy site in the $D_{3d}$ symmetry~\cite{Jones1996_limitations}. 

Hydrogen impurities in P-doped diamond passivate the $n$-type behaviour~\cite{Wang2002}, highlighting the importance of control of the growth conditions.
However, computational studies have also predicted~\cite{Czelej2018_clustering} that hydrogen in a P$V$ complex can lead to the formation of electrically, magnetically or optically active centers.
Finally, co-doping with N has been shown~\cite{Nadolinny2011_transformation} to lead to the formation of NP impurity complexes during high temperature annealing. 


\subsubsection{As}

Substitutional arsenic in diamond has been predicted to be an effective $n$-type dopant~\cite{Sun2020,Sque2004}.
In the substitutional configuration the As remains in the tetrahedral $T_d$ site pushing its nearest neighbors away and causing the next C-C bonds to slightly compress.
The formation energies indicate that incorporation of individual As is possible.
A single substitutional As introduces an impurity level at the lowest level of diamond conduction band reducing the band gap.
The ionization energy of the single dopant is reported~\cite{Sun2020} at $0.23$~eV, indicating that the structure is a promising system for creating $n$-type diamond.
Another study~\cite{Sque2004} shows a donor activation energy of $0.4$~eV for substitutional As, confirming the shallow donor.

On the other hand, it has been shown that vacancy-type defects in the lattice greatly affect doping, acting as compensating centers.
This behaviour highlights the importance of controlling the additional defects in the lattice.
No clear indication of spin-polarization was seen in the doped structures.
With a neighboring vacancy, the substitutional As will take the split-vacancy configuration and is expected~\cite{Goss2005} to be optically active.
All studied As complexes with hydrogen show only deep donor levels~\cite{Sque2004}.

The incorporation of As impurities into diamond during growth has not been successful in a hot filament reactor~\cite{May2007}.
However, Ref.~\cite{Zaitsev2001_handbook} cites reports on As impurities that have been included in diamond during growth.
The reported optical bands are $695$, $2448$, $3534$ and $4130$~nm, likely linked to the impurity centers, as well as the $2365$~nm band accompanied with two lines at $2145$ and $2883$~nm.

\subsubsection{Sb}

Research on antimony impurities in diamond mainly consists of theoretical work.
Substitutional Sb has been found~\cite{Sque2004} to energetically favor the tetrahedral $T_d$ site pushing the nearest C atoms back significantly in pristine diamond lattice.
Sb is a shallow donor.  
Sb substitution with a neighboring vacancy results in the split-vacancy configuration with the $D_{3d}$ symmetry~\cite{Goss2005}.
In the case of hydrogen co-impurities in the lattice, the SbH complex has been shown to induce a deep donor level in the band gap~\cite{Sque2004}.

\subsection{Group 16: chalcogens}

Group $16$ elements have been studied as likely donors in diamond.
Beyond introducing gap states, sulfur has been found to be an optically active center with potential for spin manipulation. 

\begin{table}[h!]
    \caption{
    {\bf Group 16.} 
    States introduced in the diamond band gap by the group $16$ impurity centers are indicated by the nature of the doping: acceptor ($p$-type) and donor ($n$-type). The optical zero phonon transitions are indicated when available.
    }
    \begin{tabular}{ |c|c|c| }
    \hline
    impurity & gap state & zero phonon line (nm)\\
    \hline
    O$_s$ & donor & 498.2, 502.8, 584.8,\\
    && 598.3, 836, 845\\
    \hline
    S$_s$ & donor/deep donor & -\\
    S$V$ & acceptor & 1107, 1016\\
    SN & donor &-\\
    SB & donor &-\\
    \hline
    Se$_s$ & deep donor & -\\
    SeB & shallow donor &- \\
    SbH & shallow donor &-\\
    \hline
    Te$_s$ & shallow donor&- \\
    TeH & donor &-\\
    \hline 
    \end{tabular}\\
    \footnotesize \noindent X$V$: X plus vacancy, XH: X with hydrogen, XB: X with boron, X$_s$: X in substitutional site
    \label{table::group_16}
\end{table}

\subsubsection{O}

Oxygen doping can introduce donor levels in the diamond band gap. 
It is likely to take a substitutional configuration in diamond as was predicted theoretically already in 1996~\cite{Anderson1996}.  
The substitutional O relaxes into the site without lattice distortions~\cite{Ullah2015}.
It is a thermodynamically favorable doping process due to the negative formation energy of a substitutional oxygen.
Because of the different electronegativities of carbon and oxygen, the C-O bond can be polarized driving the shift in the electronic properties of oxygen-doped diamond.
Charge density shows that the C-O bond is chemically stronger than a C-C bond, although also opposing results have been published~\cite{Gali2001}, indicating a weaker C-O bond due to a significant outward relaxation around the O creating a highly localised distortion.
Earlier studies~\cite{Goss2005} also report on asymmetric energy minimum configurations.

Oxygen doping creates two new impurity levels in the diamond band gap~\cite{Ullah2015,Gali2001,Barnard2003,Goss2005}.
When a level is created close to the conduction band minima, the Fermi level drops below the conduction band minima and the substitutional oxygen acts as a donor~\cite{Ullah2015}.
The diamond band gap is decreased, depending on the amount of dopants, hence increasing the electrical conductivity and making oxygen-doped diamond act as a semiconductor.

Optically, oxygen has been linked~\cite{GIPPIUS1993_oxygen} to the $498.2$, $502.8$, $584.8$, $598.3$, $836$ and $845$~nm zero phonon lines in oxygen-implanted natural diamond, similar to the $267.2$ and $330.62$~nm luminescence peaks~\cite{Mori1992}.
Oxygen ion implantation into diamond has been reported~\cite{LühmannTobias2018Saeo}, where intense fluorescence of O-related centers was observed after annealing at $1600$$^\circ$C, with the zero phonon line at $584.5$~nm.

\subsubsection{S}

Sulfur has been studied as a potential $n$-type donor in diamond, but has yet to show significant promise. 
Interestingly, it has recently been proposed~\cite{CHENG2017_theory} to be an optically active spin coherent center.
Sulfur-implanted and annealed samples exhibit conductivity with an activation energy of $0.19-0.42$~eV~\cite{Hasegawa_1999_ntype,TROUPIS2002}.
Although Hall measurements did not provide consistent results~\cite{Hasegawa_1999_ntype}, a $pn$-junction was fabricated from S- and B-doped layers and the results support the conclusion of $n$-type behaviour due to the S dopants.
S impurities in CVD-grown diamond have also been linked to $n$-type doping~\cite{PSakaguchi1999_sulfur,NISHITANIGAMO2000,Nakasawa2003_cathodl}.
However, it has been suggested that accidental contamination with B could have caused the observed shallow donor behaviour~\cite{kalish2000_sulfur}.
Nevertheless, further investigations~\cite{CHENG2017_theory} support the claim that S is a donor in diamond.

Most computational results indicate~\cite{MIYAZAKI2001_theoretical,NISHIMATSU2001_theoretical,Katayama-Yoshida_2001_codoping,Wang2002,Sque2004,lombardi2004_interaction,Goss2005,cai2006_origin,Barnard2003,Czelej2017} that substitutional S is a deep donor, although a more shallow gap level has alos been reported~\cite{Anderson1996,saada2000_sulfur,Zhou_2001_quantum}, depending on the exact computational details.  
Ionisation leads to a slightly deeper energy level ($0.5$~eV below conduction band). However, the doubly charged S$^{2+}$ site, which has the lowest formation energy and thus is the most likely configuration, does not introduce any gap states~\cite{saada2000_sulfur}.
Such impurity centers could decrease the $n$-type behavior.  

Complex sites with a vacancy and co-doping configurations with N, B and even a certain configuration with a single H~\cite{lombardi2004_interaction}, lead donor levels in a more shallow range of $0.4-0.6$~eV~\cite{MIYAZAKI2001_theoretical}.
However, sites with vacancies may result in a configuration that in a negative or a neutral charge state is an acceptor~\cite{Baker2008_electonparam,CHENG2017_theory}.
Ion implantation and high temperature annealing is one way to obtain S-containing complexes. 
However, this may not produce the intended $n$-type behavior due to the subsequent formation of $p$-type split-vacancies. 

Recently~\cite{CHENG2017_theory}, the S-vacancy (S$V$) site was predicted to be an optically active center with zero phonon lines at $1107$ and $1016$~nm.
The center was also connected to a long spin coherence time, predicted to be at the similar range to the exceptional N$V^-$ center.
Due to its nature as a deep acceptor, the S$V$ center can not be thermally excited, but indicates two separate photon-driven excitations to spin-conserved triplet states.
Due to the two triplet states, the spin initialisation and detection can be achieved optically with two separate energies.
Thereby the S$V$ center may provide a more varied center for spin manipulation compared to the N$V^-$.

\subsubsection{Se}

Selenium has been studied as a potential donor in diamond.
Studies have not been able to find shallow behaviour, although this could be achieved with co-doping scenarios.
On a substitutional site Se likely takes the $C_{3v}$ symmetry~\cite{Goss2005} and introduces deep donor levels in diamond~\cite{Sque2004,Czelej2017}.
The formation energy of the Se substitution remains high ($14.476$~eV)~\cite{WU202111} due to the much larger covalent radius of Se compared to C.
With a neighboring vacancy, Se will adopt the $D_{3d}$ symmetry~\cite{Goss2005} and will compensate any shallow donor states.
Possible optical activity of the Se-vacancy complex, like that of the S-vacancy, has not been confirmed.
Co-doping with boron has been predicted to generate $n$-type shallow donor levels~\cite{WU202111} similar to the behavior when co-doping with hydrogen~\cite{Sque2004}, but the formation energies are high.

\subsubsection{Te}

Another potential shallow donor in diamond, albeit scarcely studied, substitutional tellurium produces~\cite{Goss2005} an impurity level at $0.7$~eV below the diamond conduction band gap minimum assuming the $C_{3v}$ symmetry. 
Co-doping with H~\cite{Sque2004} is likely to induce gap states at about $0.5-0.7$~eV below the conduction band minimum.

\subsection{Group 17: halogens}

Group $17$ elements could offer some potential donors in diamond, if the atoms were successfully introduced into the lattice.
However, formation energies are high, which indicates ion implantation as the only promising route for fabrication. 
Fluorine is the only element in the group that has so far shown any promise as an optically active center.

\begin{table}[h!]
    \caption{
    {\bf Group 17.} 
    States introduced in the diamond band gap by the group $17$ impurity centers are indicated by the nature of the doping: acceptor ($p$-type) and donor ($n$-type). The optical zero phonon transitions are indicated when available.
    }
    \begin{tabular}{ |c|c|c| }
    \hline
    impurity & gap state & zero phonon line (nm)\\
    \hline
    F$_s$ & acceptor  &-\\
    F$_i$ & shallow acceptor & -\\
    F (related) & - & 600 \\
    \hline
    Cl$_s$ & donor &-\\
    Cl$_i$ & deep donor &-\\
    Cl$V$ & deep donor &- \\
    ClH & deep donor &-\\
    \hline
    Br$_s$ & donor &- \\
    Br$V$ & deep donor &-\\
    BrH & deep donor &-\\
    \hline
    I$_s$ & shallow donor &- \\
    I$V$ & acceptor/shallow donor &-\\
    IH & deep donor &-\\
    \hline 
    \end{tabular}\\
        \footnotesize \noindent X$V$: X plus vacancy, XH: X with hydrogen, X$_s$: X in substitutional site, X$_i$: X in interstitial site
    \label{table::group_17}
\end{table}

\subsubsection{F}

Fluorine impurities show promise as color centers in diamond~\cite{Tchernij2020}.
Photoluminescence is reported with a weak emission line at $558$~nm and two intense lines at $670$ and $710$~nm.
Measurements at cryogenic temperatures suggest a zero phonon line at $600$~nm.
F implanted into diamond via ion implantation~\cite{LühmannTobias2018Saeo} exhibited no visible zero phonon line but a broad emission peak centered around $680$~nm. The study reports that single fluorescence spots have been determined to be single-quantum-emitters.

Calculations reveal~\cite{Yan2009} that F favors the substitutional site in the diamond lattice.
It has a negative formation energy, indicating a thermodynamically favorable doping.
The interstitial site has a significantly higher formation energy, making it less likely, although accessible via ion implantation.
A 2005 study~\cite{Goss2005} indicates a distorted F substitution in an orthorombic $C_{2v}$ symmetry.
The substitutional site introduces impurity levels in the band gap above the valence band maximum, where the Fermi level is pinned~\cite{Yan2009}.
This leads to F acting as an acceptor in diamond.
In an interstitial configuration, impurity levels are created close to the valence band maximum, indicating a $p$-type conduction and the F as a shallow acceptor.
Neutral substitutional F shows paramagnetic behavior~\cite{Goss2005}.
If future electron paramagnetic resonance experiments confirm this, the F center could present a new system for practical applications in quantum information technology. 

\subsubsection{Cl}

Chlorine is a potential donor in diamond, although the introduced gap state is likely a deep one.
A substitutional Cl impurity is likely to prefer tetragonal $C_4$ structure in diamond~\cite{Anderson1996}.
Calculations predict paramagnetic behaviour~\cite{Goss2005}.
A study~\cite{Yan2009} comparing Cl in interstitial and
substitutional sites reports on the formation energies,
which are over 6 eV in favor of the substitutional site.
On both sites the Cl donates electrons to the diamond, making it a donor, although the amount of donated charge is lower than for the other halogens Br and I.
Substitutionally Cl-doped diamond was reported~\cite{Yan2009} to have two impurity levels within the band gap, with the Fermi level located below the conduction band minimum.
However, no shallow donor behavior was confirmed, as was predicted by one of the earliest calculations~\cite{Anderson1996}.
In the interstitial site Cl induces deep impurity levels in the band gap that are located above the valence band maximum where the Fermi level also resides. In this configuration the impurity acts as a deep donor. 

With a neighboring vacancy, the Cl$V$ stays in the substitutional site ($D_{3d}$ symmetry).
A site with a neighboring vacancy is clearly favored over the simple substitution (energy difference $4$~eV)~\cite{Goss2005}.
Therefore, Cl is likely to combine with existing vacancies in the lattice forming impurity vacancy complexes.
For Cl$V$, impurity states appear above the valence band maximum where also the Fermi level is pinned.
The Fermi level is far away from the conduction band minimum, which indicates a deep donor state. 

The effect of a hydrogen impurity in the substitutional and interstitial Cl has been reported~\cite{Yan2009} to disperse the impurity bands and shift the Fermi energy downwards in the band gap.
The incorporation of H in the structure therefore reduces the donor behaviour of Cl.   

\subsubsection{Br}

Bromine is another deep donor of the group $17$ elements.
It donates electrons to the lattice and leads to a gap state below the conduction band minimum.
The substitutional configuration has a trigonal $C_{3v}$ symmetry~\cite{Goss2005} and several stable charge states ($2+$, $1+$, $0$ and $1-$)~\cite{Czelej2017}.
Due to a relatively high formation energy (in a substitutional site $4.33$~eV and as an interstitial $26.6$~eV~\cite{Yan2009}), the dopant is unlikely to be introduced during growth, but requires post growth methods such as ion implantation. 
It is energetically favorable for Br to bind with vacancy-type defects in the diamond lattice forming Br$V$ complexes.
The Br$V$ induces impurity states deeper in the band gap where the Fermi level also shifts.
Also the incorporation of a hydrogen co-impurity~\cite{Yan2009} will shift the Fermi level towards the valence band maximum increasing the activation energy of the Br donor.

\subsubsection{I}

Iodine substitution in diamond has a relatively high formation energy ($9.64$~eV), while the interstitial site is even less likely with a formation energy as high as $34.08$~eV~\cite{Yan2009}.
The substitutional I brings the Fermi level to ca. $0.4$~eV below the conduction band minimum.
This makes it a shallow donor and a promising candidate for $n$-type doping, if the large atom can be introduced in the lattice.

With a neighboring vacancy, it has been predicted~\cite{Yan2009} that the formation energy of the I substitution drops down to $-0.3$~eV, indicating a strong preference for pairing with vacancies.
With the vacancy, the impurity level is introduced near the middle of the band gap and the Fermi level crosses the top of this band.
This indicates that the vacancy complex I$V$ does not behave as a shallow donor, but as a compensating one~\cite{Goss2005}, although much earlier calculations~\cite{Anderson1996} reported it to be a shallow thermal donor.
With the tendency of I to pair with vacancies, this may hinder the application of I-doped diamond. 
Hydrogen co-doping will result in lowering the Fermi level and reducing the $n$-type behaviour~\cite{Yan2009}.

\subsection{Group 18: noble gases}

Group $18$ impurities offer a new family of optically active centers in diamond.
Although inert in gas, argon, krypton and xenon are predicted to create new states within the diamond band gap, as the tight lattice forces the impurities to react chemically with the surrounding atoms.

\begin{table}[h!]
    \caption{
    {\bf Group 18.} 
     States introduced in the diamond band gap by the group $18$ impurity centers are indicated by the nature of the doping: acceptor ($p$-type) and donor ($n$-type). The optical zero phonon transitions are indicated when available.
 }
    \begin{tabular}{ |c|c|c| }
    \hline
    impurity & gap state & zero phonon line (nm)\\
    \hline
    He (related) & no & 536.5, 560.5\\
    \hline
    Ne (related) & no & 519, 659, 658, 716,  719.5 \\
    \hline
    Ar$_i$ & no & - \\
    \hline
    Kr (related) &  - & - \\
    \hline
    Xe (related) & - & 793.3, 811.6 \\
    \hline 
    \end{tabular} \\
        \footnotesize \noindent X$_i$: X in interstitial site
    \label{table::group_18}
\end{table}

\subsubsection{He}

He is a potential optical center in diamond. 
Luminescence lines in He$^+$-implanted diamond were first observed at $513$, $522$, $535$, and $560$~nm~\cite{GIPPIUS1983}, and later multiple additional zero phonon lines were measured~\cite{Tkachev1985}.
Most recent work~\cite{FORNERIS2015,Prestopino2017} focuses on the sharp emission lines at $535.2$ and $559.7$~nm, which persist also after annealing at temperatures above 500$^\circ$C, demonstrating the stability of the He color center.
The $522.5$~nm line anneals sharply at $850^\circ$C, and the $536.5$ and $560.5$~nm lines at $1200^\circ$C~\cite{Tkachev1985}.
Of those, the zero phonon emission was tentatively assigned to the last two~\cite{Khomich2019}.

Computational work~\cite{Beck2023} discusses different configurations.
No midgap states were predicted for any of the studied sites within nanodiamond structures.
In bulk diamond the interstitial configuration with $T_d$ symmetry is favored~\cite{Goss2009_density}. This site is also the most stable one in nanodiamond~\cite{Beck2023}.
Substitutional He will cause large repulsion in the surrounding lattice and the neighboring carbon atoms relax outwards from the dopant.
The preferred geometries include $C_{2v}$, but also the $D_{2d}$ symmetry has been reported. 
The migration barrier of the He substitution was estimated to be $4.9$~eV. 
In a double vacancy the He will take the substitutional site of one of the missing carbon atoms, thus the split-vacancy is unlikely, although reportedly~\cite{Goss2009_density} only $0.2$~eV higher in energy.

\subsubsection{Ne}

Luminescence lines in Ne$^+$-implanted diamond have been observed at $513$, $535$ and $560$~nm~\cite{GIPPIUS1983}.
An additional broad band has been observed between $700$ and $850$~nm~\cite{Tkachev1985}, where the zero phonon line was assigned to a doublet at $658$ and $659$~nm as well as at $716$ and $719.5$~nm. Additionally, a possible zero phonon line linked to a Ne center has been reported~\cite{ZAITSEV1992179} at $519$~nm.
Computational work~\cite{Beck2023} compared different configurations in nanodiamond.
Similarly to He, no midgap states were predicted for Ne in nanodiamond.
The most stable site is the Ne interstitial with the $T_d$ symmetry.
In a substitutional site, Ne will push the local lattice outwards, relaxing in the $D_{2d}$ symmetry. 
Double vacancy configuration relaxes into a split-vacancy with  $D_{3d}$ symmetry and is the least stable of the studied sites.

\subsubsection{Ar}

So far, no photoluminescence has been measured in Ar-doped diamond.
Nevertheless, Ar substitution and double vacancy configurations are predicted~\cite{Beck2023} to create midgap states in the band gap of nanodiamond, that can produce transitions in the visible range of the spectrum.
This is achieved when the atom is able to chemically react with the diamond, similar to Kr and Xe.
The interstitial is the most stable of the studied configurations, but no gap states are linked to it, which might explain why they have so far not been detected experimentally.
Interstitial Ar prefers a highly distorted local geometry moving in the [111] direction inside the high symmetry tetrahedron in the lattice.
In a substitutional site, there is large repulsion between the Ar and the carbon atoms similarly as with the other noble gases. At a double vacancy site Ar will assume the split-vacancy configuration which is also the least stable of these sites. 

\subsubsection{Kr}

Krypton is predicted~\cite{Beck2023} to create midgap states in the band gap of nanodiamond with all the studied sites.
The calculated absorption spectrum features transitions in the visible range indicating similar trends for Kr as seen above for Ar.
Interstitial Kr moves in the [111] direction inside the high symmetry tetrahedron in the lattice and presents the most stable configuration.
Substitutional Kr will cause the lattice to expand locally, leading to the $D_{2d}$ symmetry.
In a double vacancy Kr will occupy the split-vacancy site. 

\subsubsection{Xe}

Xenon is optically active and predicted to introduce new states in the diamond band gap. 
Xe-implanted diamond shows a zero phonon line at $811.6$~nm~\cite{ZAITSEV1992179} and vibronic side bands.
Gradual annealing from $300^\circ$ to $1400^\circ$ shows a second zero phonon line at $793.3$~nm~\cite{MARTINOVICH2000_photol}. 
Computational work with nanodiamond~\cite{Beck2023} predicts that many Xe configurations create gap states with absorption spectra featuring transitions in the visible range.
Measurements with polarised luminescence~\cite{BERGMAN200792} indicate that the zero phonon lines originate from a [111]-oriented center, which could be a split-vacancy site.
Further measurements show~\cite{SANDSTROM2018182} that the intensity of the Xe-related emission depends on the nitrogen concentration in the diamond.
An order of magnitude higher emission was observed in a nitrogen-rich diamond where excess electrons are provided by the N, indicating a negatively charged Xe complex. 
Observing the emission from a single site has not yet been successful.
The site has a very short excited state lifetime ($0.73$~ns), but it is comparable to the Si$V$, which can be detected as an isolated site, thus not ruling this out for Xe.
So far, the Xe optical centers have been successfully used in a diamond light emitting diode~\cite{zaitsev2006_diamond} with emission at $812.5$ and $794$~nm at room temperature.

Calculations predict the Xe configurations in diamond as follows:
In an interstitial site Xe moves along the (111) plane from the center of the tetrahedron formed by the carbon atoms.
In a substitutional site the Xe will cause the lattice to expand locally due to large repulsion, leading to the $D_{2d}$ symmetry~\cite{Beck2023}, similar to the other noble gases.
In a double vacancy the atom occupies the split-vacancy site~\cite{Goss2009_density,Drumm2010,Beck2023} in a $D_{3d}$ symmetry, although also the off-center configuration was reported early on~\cite{Anderson1996}.
Interestingly, both the substitutional site~\cite{Beck2023} and the split-vacancy site~\cite{Drumm2010} have been reported as the most stable one for Xe, most likely due to different supercell structures used in the studies.
In a vacancy cluster with three vacancies, Xe takes an interstitial position between the vacancy sites.
Noteworthy is that energetically this site is less stable (by $2.59$~eV) than the split-vacancy site~\cite{Drumm2010}, which has less space to accommodate the large impurity.


\subsection{Lanthanides: rare earth elements}

Lanthanides present the most recent addition to the list of promising impurity centers in diamond. Especially interesting are europeum and erbium as optically active centers.

\begin{table}[h!]
    \caption{
    {\bf Lanthanides.} 
    The zero phonon lines for the lanthanide impurity centers.}
    \begin{tabular}{ |c|c| }
    \hline
    impurity &  zero phonon line (nm) \\
    \hline
    Pr$V$ &  1217.5\\
    \hline
    Eu (related) & 502 \\
    \hline
    Er (related) & -\\
    \hline
    \end{tabular}\\
        \footnotesize \noindent X$V$: X plus vacancy
    \footnotesize \noindent 
    \label{table::group_lanth}
\end{table}

\subsubsection{Pr}

Simulations predict~\cite{Xin2018} the stability of praseodymium-containing vacancy centers in diamond. 
Structure relaxation predicts a split-vacancy configuration, and the calculated electronic band structure the formation of spin polarized gap states corresponding to a zero phonon line of $1217.5$~nm. 

\subsubsection{Eu}

Europium is one of the recently discovered optically active impurity centers in diamond.
It has been successfully introduced into diamond via different growth processes~\cite{Magyar2014,PALYANOV20218,VANPOUCKE2019,Yudina2022,Lebedev2023}. 
It shows strong photoluminescence effect measured at about $612$~\cite{Magyar2014} and $502$~nm~\cite{PALYANOV20218}.
The transitions have been linked to the positively charged Eu$^{3+}$ ion~\cite{Yudina2022}.
Calculations confirm~\cite{VANPOUCKE2019} that the large Eu prefers a site with a neighboring vacancy (Eu$V$) over the pure substitutional site.
The Eu$V$ complex takes the split-vacancy configuration.
Nevertheless, the simulation results show that Er on a substitutional site is the most likely configuration to account for the photoluminescence at the visible region of the spectrum, as the Eu$V$ site is not predicted to show luminescence.
With two neighboring vacancies, the Eu could give rise to luminescence at long wavelengths near infrared. 
The calculated band structure of the Eu$V$ center indicates~\cite{Tan2020} the formation of spin-up and spin-down bands in the gap.
This gives rise to the favorable spin polarisation for single photon emitters.
The electronic transition energies between the spin-up unoccupied energy level and the occupied valence band corresponds to $623$~nm.
Nevertheless, the band structure also contains metastable energy levels that may hinder the quantum efficiency.
Any N co-dopants will also likely affect negatively the formation of the Eu-related single photon source.  

\subsubsection{Er}

Erbium is an optically active center in diamond.
Er dopants in solid matrices yield promising electron spin coherence times in the ms range~\cite{Er_spincoherence,Gupta2022} making such structures favorable for quantum information networks.
In diamond, Er-doping has been achieved via ion implantation.
Not surprisingly, the experiments show a high degree of disorder after the irradiation with the large dopant.
Luminescence was clearly evident in the samples after high temperature annealing at the near-infrared region~\cite{Cajzl-Er}.
Two luminescence peaks at $1503$ and $1535$~nm can be observed after annealing to $600^\circ$C.
In the ion irradiated sample, the erbium does not show a preference for substitutional positions, where only $5$ to $10\%$ of erbium atoms occupy them. 

However, structural optimization suggests~\cite{Cajzl-Er} that single Er prefers a substitutional site in the lattice over the interstitial one.
Presence of nearby vacancies changes the geometric structure and lowers the formation energy. 
Uneven number of vacancies yield the lowest energy configurations. 
In the interstitial site, Er prefers to have a neighboring vacancy site.
In this case the Er moves towards the vacancy, leading to a geometry very similar to the substitutional defect.
Sites with multiple neighboring vacancies are energetically less favorable.
In all cases, as expected for such a large impurity, the surrounding carbon atoms are displaced outwards from the dopant.

\section{Visibility of selected impurity atom configurations in STEM-HAADF images}

Despite the extensive available literature on impurities in diamond, so far no direct electron microscopy imaging of the individual impurities in the lattice has been reported.
Therefore, STEM-HAADF (high-angle annular dark-field) image simulations were performed for selected impurity atom configurations with promising properties for future applications.
They serve as a guide for electron microscopy, comparing different atomic species and configurations as well as crystal orientations.

The selected impurities were chosen to represent a wider class of dopants of which similarities are expected in the resulting images, such as lighter versus heavier elements and split-vacancy versus substitutional structure.
In total $13$ impurity structures were relaxed using density functional theory calculations.
For the dopant-vacancy pairs, the relaxed structures include nitrogen, boron, silicon, phosphorous, nickel, germanium, platinum, lead and erbium.
For substitutional structures, fluorine, chlorine, antimony and erbium were simulated.

\subsection{Structural optimization}
\label{sec::dft}

The impurity atom structures were modeled using the atomic simulation environment (ASE)~\cite{ase-paper}, incorporating the defects into the middle of a $3\times3\times3$ diamond super cell.
The structures were then relaxed using density functional theory (DFT) calculations as implemented in the GPAW~\cite{gpaw-paper} code, based on the projector augmented wave method.
The PBE~\cite{PBE-paper} exchange functional  was used.
The structures were first relaxed using the linear combination of atomic orbitals (LCAO) method with the double zeta plus polarization (DZP) basis set until all forces were below $0.05$~eV/Å to speed up the calculation, and then relaxed again using the finite difference method with a real space grid spacing of $0.2$~Å and Monkhorst-Pack $k$-point sampling of $2\times2\times2$, until all forces were below $0.01$~eV/Å.
The relaxed atomic structures for substitutional chlorine and antimony as well as the nitrogen-vacancy and erbium-vacancy were shown above in Fig.~\ref{fg:dft}.

\subsection{Image simulations}

STEM-HAADF images were simulated based on the DFT-relaxed structures using \textit{ab}TEM~\cite{abtem}.
A field of view of $6\times6$~Å$^2$, HAADF detector angles between $80-200$~mrad and convergence semiangle of $30$~mrad were used.
For all simulations an electron beam energy of $100$~keV was assumed.
The potential and probe sampling were set to $0.03$~Å, and Nyquist sampling was employed for the grid scan.
The finite Lobato parametrisation was used for the potential.
An interpolation to $0.02$~Å sampling was applied to the resulting images.
The probe was modeled to be focused on the surface of the slab, and multiple slabs with either $20$ or $40$~nm thickness were created, where the relaxed cell containing the defect was seamlessly incorporated in a larger pristine lattice.
The dimensions of the relaxed cell containing the defect were $1\times1\times1$~nm$^3$, and the lateral dimensions of the full slab were $4.5\times4.5$~nm$^2$, with the relaxed cell in the (lateral) center.
The $z$-position of the dopant within the slab was varied with a step size of $1$~nm to simulate different defocus values.
Multislice simulations of these slabs were then performed.
For the effects of temperature, selected simulations were additionally carried out including the frozen phonon approximation as given in the abTEM package, with a standard deviation of $\sigma=0.04$ (value for diamond at $300$~K), averaged over $20$ configurations.

\subsubsection{Dopant-carbon contrast-defocus dependency}

\begin{figure}
\includegraphics[width=0.45\textwidth]{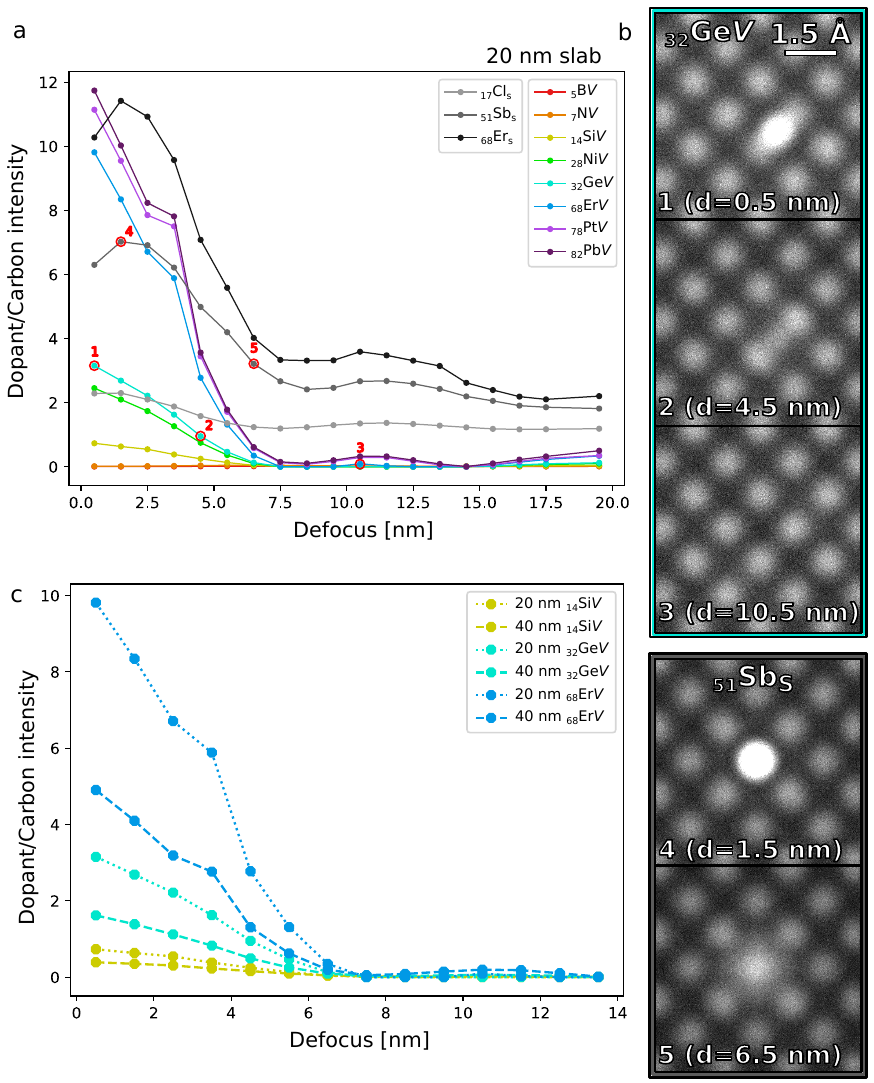}
    \caption {{\bf Dopant visibility as a function of defocus in simulated STEM-HAADF images.}
    (a) Ratio of contrast values for an atomic column containing the dopant atom and a pure-carbon column as a function of the defocus.
    (b) Simulated image examples for a dopant-vacancy configuration (Ge$V$) and a substitutional dopant (Sb$_s$) at defocus values marked in panel (a), with added Poisson noise.
    All values shown in the graphs correspond to images without the added noise.
    (c) Ratio of contrast values as a function of the defocus for selected dopant configurations in slabs with two different thicknesses ($20$ and $40$~nm).
    All simulations are for the (100) orientation. The atomic numbers are indicated in the standard notation.}
\label{fg:main_alldopants_contrast}
\end{figure}

In order to discern the dopants within diamond, a sufficient contrast difference between the column containing the dopant and the pure-carbon columns is necessary.
Unsurprisingly, the image simulations reveal that the most important factors for visibility are the position of the dopant relative to the focus of the probe and the atomic number of the dopant.
Fig.~\ref{fg:main_alldopants_contrast}a shows the contrast ratio as a function of defocus for different dopant configurations.
There is a marked difference between the results for split-vacancy configurations and those for substitutional impurities: the split-vacancy dopants become invisible in a 20-nm thick diamond at a defocus of ca.~$7.5$~nm, independent of atomic number, whereas substitutional dopants retain a significant contrast difference for all values modeled here.
This difference is also clear in the example images shown in  Fig.~\ref{fg:main_alldopants_contrast}b.

It is interesting to compare the visibility of the split-vacancy configurations to those of substitutional impurities as a function of the atomic number of the impurity.
For example, the substitutional chlorine impurity ($Z=17$) has the same peak brightness as the nickel-vacancy configuration ($Z=28$), and retains most of this contrast for all defocus values, while the silicon-vacancy configuration ($Z=14$) has a peak contrast of less than half of that of chlorine.
Based on these results, one can expect that performing a defocus series through the diamond sample should bring especially dopant vacancy configurations in and out of view.

\subsubsection{Sample thickness and atomic number}

Until now, all results corresponded to a sample thickness of $20$~nm, which is often difficult to achieve.
Therefore, we repeated some of the image simulations for a slab with twice the thickness ($40$~nm), including examples of impurities with atomic numbers ranging from 14 (Si) to 68 (Er) in the split-vacancy configuration.
The results are shown in Fig.~\ref{fg:main_alldopants_contrast}c.
It is apparent that a thicker sample reduces the contrast and visibility, as there is more signal from the carbon atoms.
In each case, the change in the contrast is approximately one half for the 40-nm thick slab as compared to the 20-nm one.
Interestingly, the defocus value where the contrast disappears barely depends on the thickness.

\begin{figure}
\includegraphics[width=0.45\textwidth]{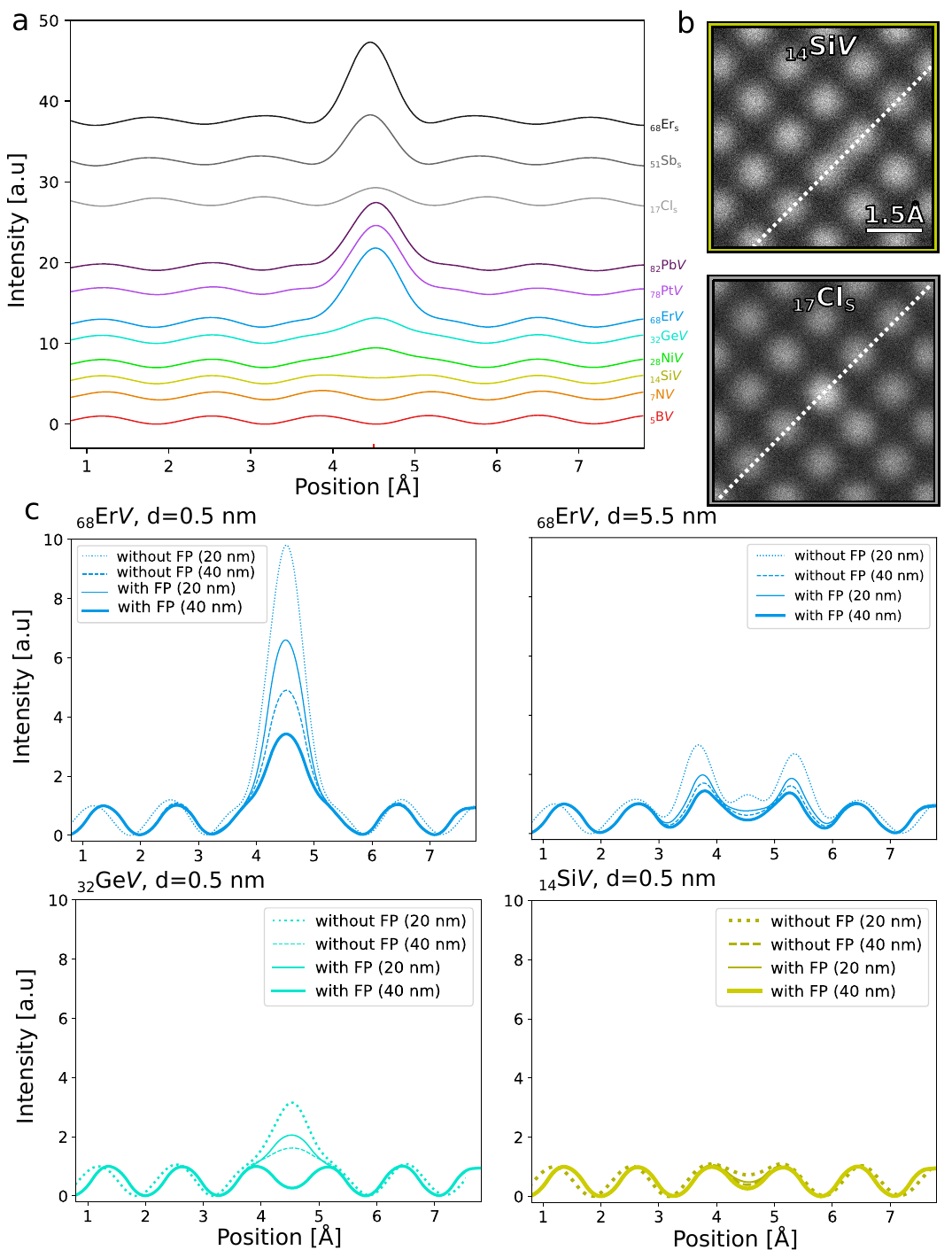}
    \caption {{\bf Intensity line profiles for simulated STEM-HAADF images for different impurity atom configurations with the optimal focus.}
    (a) Lineprofiles at $0.5$~nm defocus. Note that the intensity values have been shifted to separate the plots from each other.
    (b) Examples of simulated images for the Si$V$ and Cl$_s$ configurations.
    The white dotted lines indicate the positions of the lineprofiles.
    (c)~Intensity line profiles for images simulated with and without the frozen phonon approximation, and for two different slab thicknesses.
    The impurity configurations and the defocus values $d$ are noted in each plot.
    All simulations are for the (100) orientation. The atomic numbers are indicated in the standard notation.}
\label{fg:main_lineprofiles}
\end{figure}

The general dependence of the visibility on the atomic number can also be seen in Fig.~\ref{fg:main_lineprofiles}a, where intensity line profiles are plotted for the different impurity configurations at the optimal defocus ($0.5$~nm).
The line profile data correspond to the positions shown in Fig.~\ref{fg:main_lineprofiles}b for the Si$V$ and Cl$_s$ configurations.
It is worth noting, that for elements with atomic numbers below ca. $32$ (Ge), imaging the split-vacancy configuration becomes nearly impossible due to the low contrast difference as compared to pure-carbon atomic columns.

\subsubsection{The effect of thermal diffuse scattering}

\begin{figure*}
\includegraphics[width=.95\textwidth]{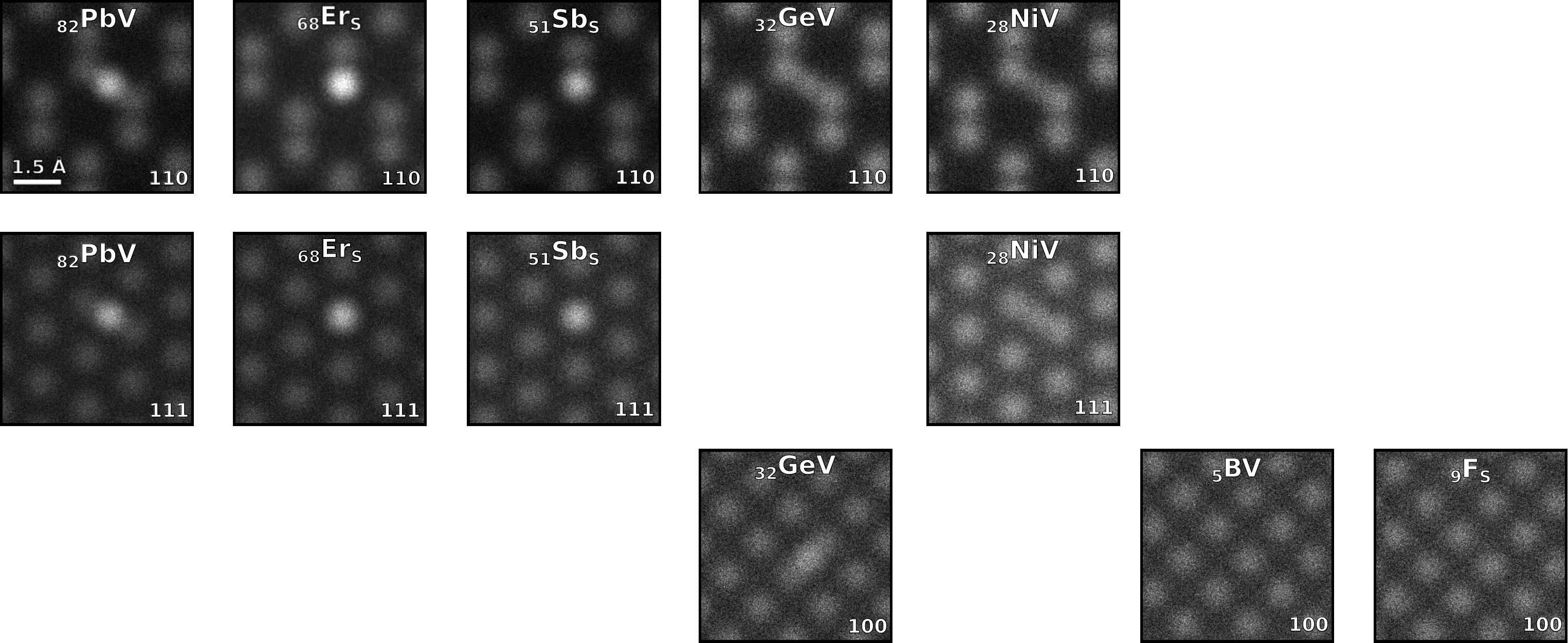}
    \caption {{\bf STEM-HAADF image simulations including the frozen phonon approximation for dopants along different crystallographic directions.}
    All simulations are made for $20$~nm slab with the optimal defocus.
    The impurity atom configuration and the crystallographic direction are noted on each image. The atomic numbers are indicated in the standard notation}
\label{fg:images_fp}
\end{figure*}

In addition to the atomic number, impurity configuration, and the slab thickness, also lattice phonons have a significant influence on STEM-HAADF image contrast, especially for thicker samples.
Due to the significant additional computational cost, we have included their effect on the image contrast only for some exemplary cases using the frozen phonon approximation (assuming $T=300$~K).
In Fig.~\ref{fg:main_lineprofiles}c, a comparison between simulations with and without the frozen phonon approximation is shown for slab thicknesses of $20$~nm and $40$~nm.
It is obvious, that the frozen phonon approximation significantly changes the visibility of the impurity configuration.
For example, in the case of Er, the dopant peak intensity drops by ca. one half when the frozen phonon approximation is included.
For many heavier impurities there is a clear increase in the intensity of the neighboring pure-carbon column for increasing defocus values. 
This effect however significantly reduces when the frozen phonon approximation is included.
Finally, for the Ge$V$ configuration, including the frozen phonon approximation completely quenches the intensity to below detection even at the optimum defocus for the $40$~nm slab, whereas for the $20$~nm slab it may still be visible.
For Si$V$, no visibility is expected.
Additional examples of image simulations including the frozen phonon approximation are shown in Fig.~\ref{fg:images_fp} (including Pb, Sb, Er, Ge, Ni, B and F). A contrast dependence on the orientation can be observed: the [110] direction should be avoided for split-vacancies, as it shows reduced contrast in comparison to the [111] and [100] directions.
For substitutional impurities there is little difference in the contrast for the three crystallographic orientations.

\subsubsection{Window of visibility}

Taking the previous results into consideration, a window of visibility (WOV) for the dopants can be estimated, yielding the range of defocus values resulting in a sufficient visibility.
Assuming that an intensity ratio of at least $1.5$ is needed for a minimal sufficient visibility in experimental conditions, and considering that the thermal effects and doubling the sample thickness reduces the intensity ratio almost by half each, the following WOV can be estimated from Fig.~\ref{fg:main_alldopants_contrast}.a: nitrogen, boron, silicon, chlorine, and nickel are not visible at any defocus value.
The Ge$V$ configuration has a very narrow WOV of $\pm1$~nm in a $20$~nm slab, while in a $40$~nm slab it is not visible.
The Er$V$ configuration has a window of visibility of $\pm4.5$~nm in a $20$~nm slab and $\pm3.5$~nm in a $40$~nm slab.
Pt$V$ and Pb$V$ configurations have a WOV of $\pm5$~nm in $20$~nm slabs and $\pm4$~nm in $40$~nm slabs.
For the Sb substitution, the WOV is $\pm7.5$~nm in a $20$~nm slab and $\pm3.5$~nm in a $40$~nm slab.
The WOV for the Er substitution is $\pm13.5$~nm in a $20$~nm slab and $\pm5.5$~nm in a $40$~nm slab. 

\subsection{Experimental visibility}

After establishing the expected visibility of impurity atoms via image simulations, we present the first experimental images of individual impurity atom configurations embedded in diamond.
A Z-cut single crystal diamond (Element Six Technologies Ltd, UK) was implanted with $190$~keV erbium ions, with an implantation fluence of $1\times10^{14}$~ions/cm$^{2}$.
The sample was then first annealed at $400^\circ$C in air, and then at $600^\circ$C in vacuum, and finally at $800^\circ$C in vacuum, each time for one hour.
A protective layer of ca. $100$~nm of gold was evaporated onto the surface followed by approximately an equal thickness of amorphous carbon to protect the sample surface from damage during further steps.
A thin vertical piece of the implanted diamond (lamella) was prepared by focused ion beam milling with a FEI Quanta 3D FEG FIB-SEM instrument employing a Ga ion beam.
The cut orientation was chosen to produce a (110) viewing orientation for electron microscopy.
The lamella was lifted out \textit{in situ} with an OmniprobeTM $100.7$ micromanipulator and attached to a copper grid where it was thinned down to its final thickness of approximately $50$~nm.
The lamella was imaged with Nion UltraSTEM 100, operated at $100$~kV, with a HAADF detector with an annular range of $80-200$~mrad, and a convergence semiangle of $30$~mrad.

\begin{figure}
\includegraphics[width=0.45\textwidth]{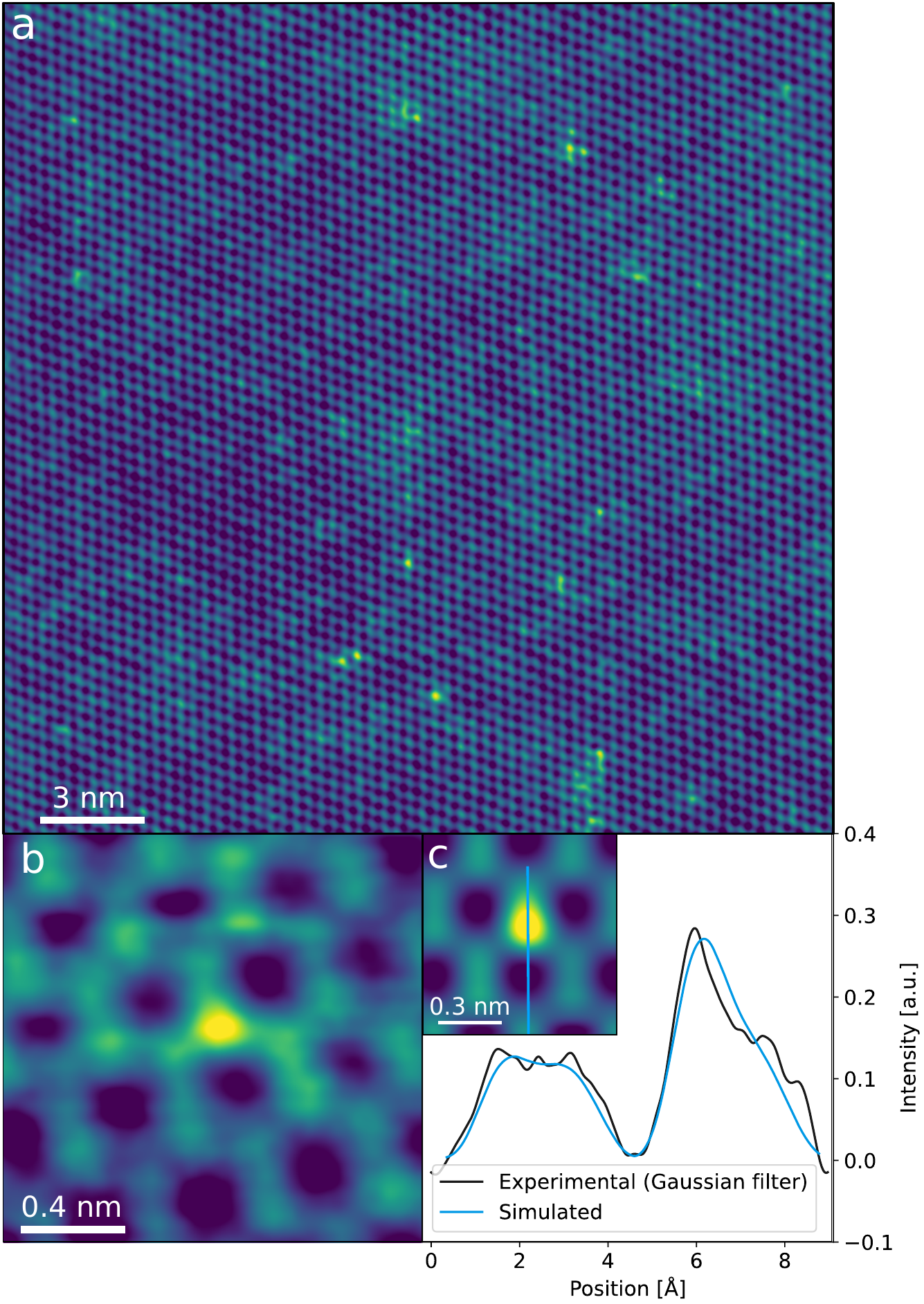}
    \caption{{\bf Er impurities in $(110)$ diamond.} (a) A STEM-HAADF image ($100$~kV beam energy) of Er-implanted $(110)$ diamond.
    Er dopants appear as bright points against the pure-carbon atomic columns.
    (b) A STEM-HAADF image of a single Er dopant (with a $6$-pixel Gaussian filter applied). (c) Simulated STEM-HAADF image of an Er substitution in diamond with a lineprofile comparison to the experimental data from panel (b) (same filter parameters).
}
\label{fg::erbium}
\end{figure} 

In Fig.~\ref{fg::erbium}, atomic-resolution STEM-HAADF images of the $(110)$-oriented diamond lamella can be seen.
Bright impurity atoms are clearly visible within the carbon lattice, as expected based on the image simulations.
The varying contrast visible at the larger scale (Fig.~\ref{fg::erbium}a) arises from possibly varying sample thickness and amorphous surface contamination.
The bright contrast impurities were only found at a depth of about $45$~nm in the lamella cross section, corresponding to the implantation depth of erbium ions as reported by J. Cajzl et al.\cite{Cajzl-Er}, with the same sample type and experimental irradiation parameters.
Image series recorded with various defoci confirm the predicted behavior of dopants fading in and out of visibility, with a WOV of roughly $\pm5$~nm.

We note that the dopants exhibit dynamic behavior under continuous imaging conditions. The individual Er atoms are seen jumping between neighboring lattice sites. We believe this is possible due to the high amount of disorder in the lattice, namely neighboring vacancy sites, that is caused by the ion implantation process with the heavy dopant.

The experimental close-up STEM-HAADF image (Fig.~\ref{fg::erbium}b) is compared to a simulated image in Fig.~\ref{fg::erbium}c.
For the simulated image, a slab of $50$~nm thickness was used with the erbium atom close to the surface (at ca. $5$~Å).
The simulation parameters were: Gaussian spread of $4$~Å, angular spread of $0.8$~mrad, defocus of $10$~Å, astigmatism of $20$~Å ($\phi=0$), coma of $200$~nm ($\phi=0$) and beam tilt of $15$~mrad in the $x$-axis and $0$ in the $y$-axis.
Frozen phonon approximation was included to simulate the effects of temperature ($300$~K).
Due to significant scan distortions in the image, which could not be considered in the simulation, the calibration of the $x$-axis in the lineprofile of the experimental data had to be adjusted with a scaling of $20\%$.
The match in proportion of the carbon peak and the erbium peak intensity is apparent, however slight disparities can be seen in the shape of the erbium peak and its overlapping carbon neighbor, as well as disparities in the lowest intensity regions between the two dumbbell peaks and at the edges of the line profiles.
This can be attributed to the fact that we do not have full knowledge of the atomic structure in the imaged region.
Especially, lattice disorder caused by implantation cannot be easily deduced and quantified.
However, the dopant is clearly present, as can be seen also from intensity variations in the carbon columns in Fig.~\ref{fg::erbium}b and 
the agreement between the experimental and simulated image is very good.
Combined with the fact that the impurity atoms were only found at the expected depth confirmed by Rutherford backscattering measurements in the previous study~\cite{Cajzl-Er}, we can confirm with relative certainty, that the observed bright contrasts indeed arise from individual Er impurity atoms within the  lattice.

\subsection{Conclusions of the provided simulations and experiments}

To motivate the electron microscopy community to join the effort to fully characterize impurity atom configurations in diamond, including the detailed atomic structure, we have provided a brief overview on the experimental and computational results on the most interesting impurity centers.
We performed density functional theory structural relaxation calculations and image simulations to estimate the visibility of some example impurity structures in STEM-HAADF images.
We specifically addressed the effects of atomic number, impurity configuration, sample thickness, defocus and thermal diffuse scattering on the visibility, assuming a $100$~kV electron beam.
The substitutional configuration was found to increase the visibility as compared to the split-vacancy configuration.
In all split-vacancy cases, the visibility falls to zero at $7.5$~nm defocus.
Taking all effects into consideration, the results imply that the elements with atomic number of $Z>50$ provide the easiest target for imaging with a visibility window of up to $> 10$~nm of defocus.
We also report on the first results of STEM-HAADF imaging on erbium-implanted diamond, which confirm the results predicted in the simulations.

\section{Acknowledgments}

The authors thank Viera Skákalová 
and Marian Varga 
for providing the Er-implanted diamond, as well as Gerlinde Habler 
for preparing the FIB lamella. 
The authors acknowledge funding 
from Austrian Science Fund (FWF) through project P34797-N36.  
Computational resources from the 
Vienna Scientific Cluster (VSC) 
are gratefully acknowledged.

\bibliographystyle{iopart-num}
\bibliography{refs}

\end{document}